\documentclass[openany]{aastex63}

\received{June 1, 2019}
\revised{January 10, 2019}
\accepted{\today}
\submitjournal{AJ} % Submitted to aas journals ?

\shorttitle{22 GHz Water Masers in Serpens South}
\shortauthors{Ortiz-Le\'on et al.}
\usepackage{graphicx}
\usepackage{grffile}
\usepackage{afterpage}
\usepackage{multirow}
\usepackage{dcolumn}
\usepackage[T1]{fontenc}
\usepackage{subfigure}

\begin{document}

\title{Discovery of 22~GHz Water Masers in the Serpens South Region}

\correspondingauthor{Gisela N.\ Ortiz-Le\'on}
\email{gortiz@mpifr-bonn.mpg.de}

\author[0000-0002-2863-676X]{Gisela N.\ Ortiz-Le\'on}
\affiliation{Max Planck Institut f\"ur Radioastronomie, Auf dem H\"ugel 69,  D-53121 Bonn, Germany}

\author[0000-0002-9912-5705]{Adele Plunkett}
\affiliation{National Radio Astronomy Observatory (NRAO), 520 Edgemont Road, Charlottesville, VA 22903, USA.}

\author[0000-0002-5635-3345]{Laurent Loinard}
\affiliation{Instituto de Radioastronom\'ia y Astrof\'isica (IRyA), Universidad Nacional Aut\'onoma de M\'exico Morelia, 58089, Mexico}

\author[0000-0001-6010-6200]{Sergio A.\ Dzib}
\affiliation{Max Planck Institut f\"ur Radioastronomie, Auf dem H\"ugel 69,  D-53121 Bonn, Germany}

\author[0000-0001-8286-2795]{Carolina B.\ Rodr\'iguez-Garza}
\affiliation{Tecnologico de Monterrey, Escuela de Ingenier\'ia y Ciencias, Ave. Eugenio Garza Sada 2501, Monterrey,  64849, Mexico}

\author[0000-0003-2133-4862]{Thushara Pillai}
\affiliation{Institute for Astrophysical Research, Boston University, Boston, MA 02215, USA}

\author[0000-0002-3866-414X]{Yan Gong}
\affiliation{Max Planck Institut f\"ur Radioastronomie, Auf dem H\"ugel 69,  D-53121 Bonn, Germany}

\author[0000-0003-4468-761X]{Andreas Brunthaler}
\affiliation{Max Planck Institut f\"ur Radioastronomie, Auf dem H\"ugel 69,  D-53121 Bonn, Germany}

%% Note that the \and command from previous versions of AASTeX is now
%% depreciated in this version as it is no longer necessary. AASTeX 
%% automatically takes care of all commas and "and"s between authors names.

%% AASTeX 6.3 has the new \collaboration and \nocollaboration commands to
%% provide the collaboration status of a group of authors. These commands 
%% can be used either before or after the list of corresponding authors. The
%% argument for \collaboration is the collaboration identifier. Authors are
%% encouraged to surround collaboration identifiers with ()s. The 
%% \nocollaboration command takes no argument and exists to indicate that
%% the nearby authors are not part of surrounding collaborations.

%% Mark off the abstract in the ``abstract'' environment. 
\begin{abstract}

Using the Karl G. Jansky Very Large Array (VLA), we have conducted a survey for 22~GHz, $6_{1,6}$--$5_{2,3}$ H$_2$O masers toward the Serpens South region.
 The masers were also observed with the Very Long Baseline Array (VLBA)  following the VLA detections. We detect for the first time H$_2$O masers in the Serpens South region that are found to be associated to three Class 0--Class I objects, including the two brightest protostars in the Serpens South cluster, known as CARMA-6 and CARMA-7.  We also detect H$_2$O masers associated to a source with no outflow or jet features. We suggest that this source is most probably a background AGB star projected in the direction of Serpens South.   
 The spatial distribution of the emission spots suggest that the masers in the three Class 0--Class I objects emerge very close to the protostars and are likely excited in shocks driven by the interaction between a protostellar jet and  the circumstellar material. Based on the comparison of the distributions of bolometric luminosity of sources hosting 22 GHz H$_2$O masers and 162 YSOs covered by our observations, we identify a limit of $L_{\rm Bol}\approx10 L_\odot$ for a source to host water masers. However, the maser emission shows strong variability in both intensity and velocity spread, and therefore masers associated to lower-luminosity sources may have been missed by our observations. We also report 11 new sources with radio continuum emission at 22~GHz. 

\end{abstract}

%% Keywords should appear after the \end{abstract} command. 
%% See the online documentation for the full list of available subject
%% keywords and the rules for their use.

%% From the front matter, we move on to the body of the paper.
%% Sections are demarcated by \section and \subsection, respectively.
%% Observe the use of the LaTeX \label
%% command after the \subsection to give a symbolic KEY to the
%% subsection for cross-referencing in a \ref command.
%% You can use LaTeX's \ref and \label commands to keep track of
%% cross-references to sections, equations, tables, and figures.
%% That way, if you change the order of any elements, LaTeX will
%% automatically renumber them.
%%
%% We recommend that authors also use the natbib \citep
%% and \citet commands to identify citations.  The citations are
%% tied to the reference list via symbolic KEYs. The KEY corresponds
%% to the KEY in the \bibitem in the reference list below. 

\section{Introduction}\label{sec:intro}

Water masers are known to be abundant in low- and high-mass star-forming regions, where they trace collimated  
outflows \citep[e.g.,][]{Furuya1999,Furuya2000,Hollenbach2013,Moscadelli2013}, and protoplanetary disks \citep{Fiebig1996,Torrelles1998}, both of which are key features during the earliest phases of protostellar evolution. 
In particular, the water maser line from the $J=6_{1,6} - 5_{2,3}$ rotational transition at 22~GHz has been detected, since its discovery by \cite{Cheung1969}, in hundreds of sources within both high- and low-mass star forming regions \citep[e.g.,][]{Furuya2003,Moscadelli2020}. These masers are extremely bright and compact, and have become primary targets for Very Long Baseline Interferometry (VLBI), which can probe angular resolutions better than 1~mas \citep[e.g.,][]{Wu2014,Sanna2017}. Observations of  22~GHz water masers have been crucial primarily for the study of the dense gas and their dynamics around young stellar objects (YSOs; \citealt{Moscadelli2019}).  

The earliest phase of low-mass protostellar evolution (the Class 0 phase in the evolutionary classes defined by \citealt{Lada1987} and \citealt{Andre1993}) is characterized by the presence of powerful outflows, which are believed to be intimately linked to the accretion process. These outflows can create shocked regions where the protostellar jets impact the ambient molecular cloud, which could  collisionally pump H$_2$O maser emission. Searches for water masers frequently target the youngest low-mass protostars, since they exhibit the most powerful collimated mass outflows. Several systematic surveys to search for water maser emission toward low-mass stars have been conducted in the past \citep{Wilking1994,Claussen1996,Furuya2001,Furuya2003}. These surveys have found that the detection rate of water masers drops drastically as protostars evolve through the Class I and II phases \citep{Furuya2001,Furuya2003}.
This is explained by the dissipation of  the dense gas around the central object as it evolves.  Also, the detection rate of water masers does seem to drop significantly for very low-luminosity objects ($L\lesssim0.4~L_\odot$; \citealt{Gomez2017}).

Only a few VLBI studies of the kinematics of water masers in low-mass stars have been conducted in the past \citep[e.g.,][]{Claussen1998,Moscadelli2006,Imai2007,Dzib2018}, in part because they are weaker than their counterparts associated to high-mass stars. These studies have shown that the masers emerge at the base of the protostellar jet, in shocks likely driven by the interaction with the disk, or in shocked gas clumps along the axis of the jet \citep{Moscadelli2006}.

\begin{figure*}
\begin{center}
 {\includegraphics[width=0.85\textwidth,angle=0]{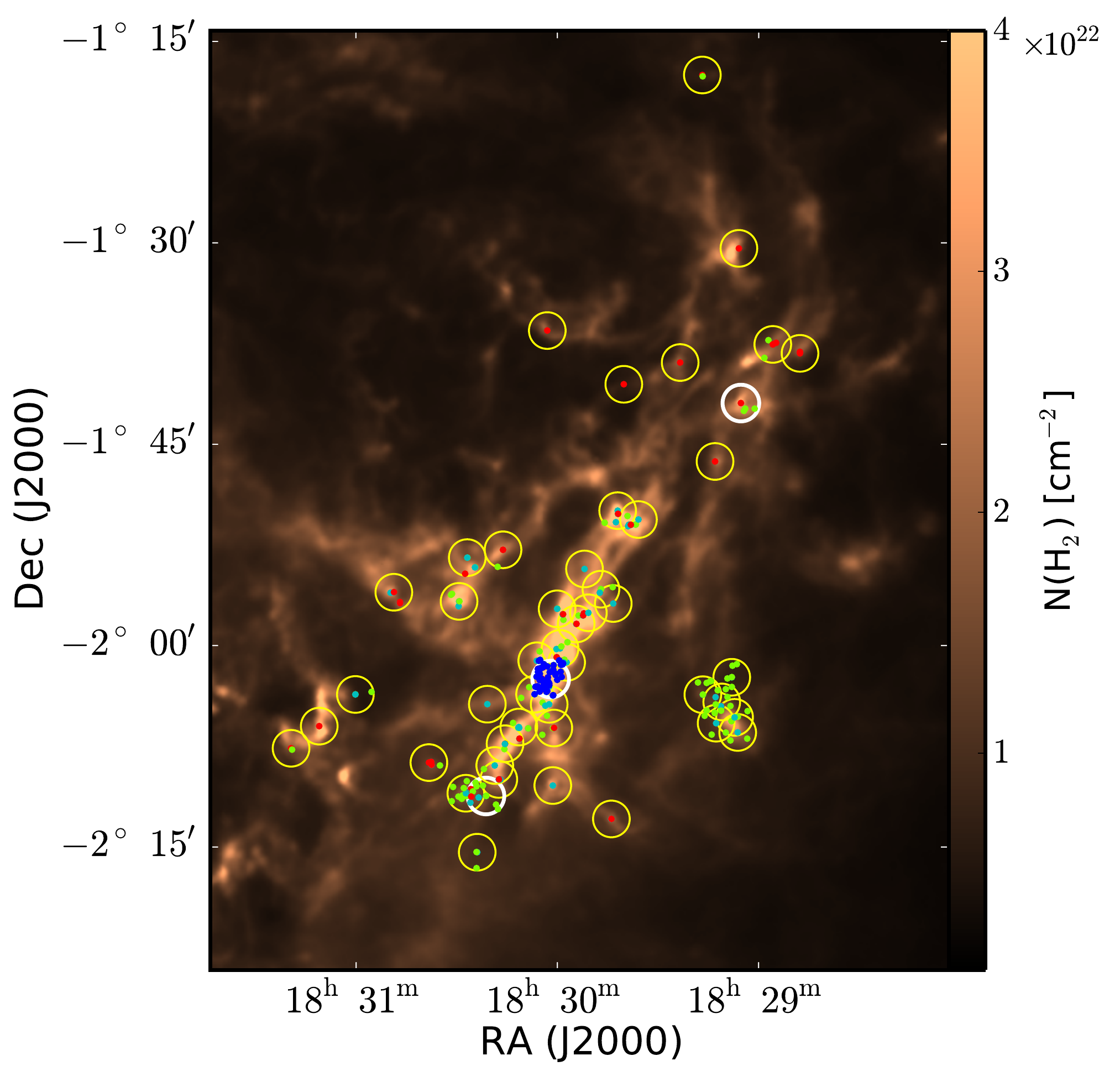}}
 \end{center}
\caption{The 48 VLA pointings used for our observations are indicated by the large circles. Fields where H$_2$O masers are detected are in white.  The circle diameters of $2\rlap.{'}7$ correspond to the field of view at 22.2~GHz. Cyan dots mark the positions of 90 young stars reported in \cite{Winston2018AJ}, which were identified as Class 0+I objects. Red dots correspond to 60 YSO candidates by \cite{Dunham_2015} classified as Class 0+I objects. Blue dots are 67 protostars identified by \cite{Plunkett2018} from ALMA 1-mm continuum observations and IR data. The distribution of the VLA pointings was chosen to cover the most of these objects. Green dots correspond to 31 Flat-spectrum, 59 Class II and 12 Class III objects from the catalog of \cite{Dunham_2015} that fall within the observed VLA fields. The background is a {\it Herschel} H$_{2}$ column density map of the Serpens South star-forming region \citep{Andre2010}. 
}
\label{fig:vla}
\end{figure*}

In this paper, we focus on the Serpens South region, a well known region harboring one of nearest very young protostellar  clusters. Serpens South was discovered by \cite{Gutermuth_2008} from {\it Spitzer} images as an infrared dark cloud and since then it has become an interesting target to observe low-mass young stars in the earliest phases of its development. 
It is located $\sim$3$^{\rm o}$ south of the Serpens Main cloud, a region also rich in star formation activity \citep{Eiroa_2008}. 
The W40 region, located $\sim$10~arcmin to the east of Serpens South, is a more evolved star-forming region hosting a cluster of massive stars and an HII region. Serpens South and W40 are both projected within the broader Aquila Rift complex of molecular clouds, and often are referred to as the Aquila region \citep[e.g.][]{Andre2010}.

The distance to the Aquila region has been a matter of debate in the literature. However, recent measurements does seem to converge to $\approx$440--480~pc \citep[e.g.][]{Ortiz2017,Zucker2019,Herczeg2019}.  \cite{Ortiz2017} obtained VLBI trigonometric parallaxes of radio continuum sources in Serpens Main and W40 and reported an average distance of $436.0\pm9.2$~pc. Later, \cite{Ortiz2018} analyzed {\it Gaia} parallaxes of stars in the Aquila region (two stars are projected in the outskirts of Serpens South) and in Serpens Main, delivered as part of the 2nd  data release (DR2). They found that the {\it Gaia} parallaxes from Aquila agree on average with those from Serpens Main, and  are  also consistent with the previous VLBI estimation, although their associated uncertainties are larger. 
Thus,  in the present study we adopt the distance from the VLBI measurement of $436.0\pm9.2$~pc.

 \begin{table*}
\caption{VLA observed epochs}
\label{tab:obs} 
\centering 
\begin{tabular}{c c c c c c c}  
\hline\hline  
Epoch & Observation  & VLA               & \multicolumn{3}{c}{Continuum}    & Channel             \\
           &      date          &configuration  & Beam  size   &  P.A. & rms   &  rms  \\
           &                       &                      & ($''\times''$) &  ($^{\rm o}$)   & ($\mu$Jy~beam$^{-1}$) & (mJy~beam$^{-1}$) \\
\hline 
%%%%%
1 &  2019 Jan 19 &  C & 1.08$\times$0.76 & $-12.5$ & 25 & 16 \\
2 & 2019 Jan 26 &  C & 1.27$\times$0.80 & $-28.7$ & 24  & 16 \\
3 & 2019 Feb 02 & C & 1.16$\times$0.78 & $-22.8$  & 27  & 21 \\
4 & 2019 Feb 08 & C$\rightarrow$B  & 0.75$\times$0.44 & $-1.46$ & 21 &18 \\
\hline 
\end{tabular}
\end{table*}

Here we use the Karl G.\ Jansky Very Large Array (VLA) to conduct a survey of water masers toward Serpens South covering the region with the highest density of protostellar objects.  
Follow up observations of the VLA-detected water masers were obtained with the Very Long Baseline Array (VLBA). The paper is organized as follows. Section \ref{sec:obs} describes the target selection, acquired observations, and data reduction. In Sect. \ref{sec:detections} we present our results and discuss the properties of the detected H$_2$O emission, the spatial and velocity distribution of maser spots, and the association of the masers with outflow activity. This section also reports the sources detected with radio continuum emission. In Sect. \ref{sec:discussion} we discuss the relationship between H$_2$O maser emission and bolometric source luminosity. Finally, Sect. \ref{sec:conclusions} presents our conclusions.

\section{Observations and data reduction}\label{sec:obs}

\subsection{VLA observations}\label{sec:vlaobs}

We observed the $6_{1,6} - 5_{2,3}$ H$_2$O maser line (at rest frequency 22,235.080 MHz) with the K-band receiver at  a velocity resolution of 0.1~km~s$^{-1}$ (corresponding to 7.8~kHz) and a velocity coverage of $\sim$100~km~s$^{-1}$. The observations were taken in 4 epochs in C and C$\rightarrow$B\footnote{C$\rightarrow$B denotes the reconfiguration from C- to B-array.} configurations (Table \ref{tab:obs}) under program 18B-230.  The epoch observed on February 2, 2019 missed 25\% of the scans; therefore it was re-observed on February 8, 2019 with the C$\rightarrow$B configuration (Table \ref{tab:obs}). The water maser line was covered by a 16-MHz wide spectral window with 2048 channels. Eight additional 128-MHz wide spectral windows (with 64 channels each) were observed in each baseband for the continuum, resulting in an aggregate bandwidth of 2~GHz.

A total of 48 VLA fields (Figure \ref{fig:vla}) were selected to cover essentially all known low-mass protostars across the region. Our target sample includes all Class 0+I candidates\footnote{Class 0+I  refers to objects in the Class 0 or Class I phase, that cannot be separated based on IR measurements alone. \cite{Dunham_2015} uses the IR extinction corrected spectral index, $\alpha$, with $\alpha\geq0.3$ for Class 0+I. \cite{Winston2018AJ} uses IR colors to identify (deeply) embedded protostars as Class 0+I objects. } 
reported in \citet[][90 sources]{Winston2018AJ} and \citet[][60 sources]{Dunham_2015}, 
as well as the 67 protostars identified by \cite{Plunkett2018}  from ALMA 1-mm continuum observations and infrared (IR) data. The observed area also includes 31 Flat-spectrum, 59 Class II and 12 Class III objects from the catalog of \cite{Dunham_2015}. 

Observing sessions  consisted of series of three scans on target fields (for $\sim$1.8~min each target) bracketed by phase calibrator scans of $\sim$1.4min.   The quasar 3C286 ($\alpha$(J2000) = 13:31:08.287984, $\delta$(J2000) = +30:30:32.95886), observed at the beginning of the observations, was used as the standard flux and bandpass calibrator, while  J1851+0035 ($\alpha$(J2000) = 18:51:46.7217, $\delta$(J2000) = +00:35:32.414) was used as the phase calibrator.  The total observing time in each epoch  was about 2.4~hr.

Data calibration was performed with the NRAO's Common Astronomy Software Applications (CASA, version 5.4.1) package using the VLA pipeline\footnote{\url{https://science.nrao.edu/facilities/vla/data-processing/pipeline/CIPL_54}} provided along with the data, that has been modified to work with spectral line observations. The calibrated visibilities were imaged using the CASA task {\tt tclean}. We produced maps of continuum emission for each observed field by integrating the full 2-GHz bandwidth. The pixel size was $0\rlap.{''}16$ in the maps from the first three epochs and $0\rlap.{''}073$ in the last epoch. The number of clean iterations was set to 10,000 with a threshold for cleaning of 0.066~mJy. We use ``Briggs'' weighting and applied the primary beam correction. For the image sizes, we used 1040$\times$1040 and 2250$\times$2250 pixels in C and C$\rightarrow$B configuration, respectively, that correspond to a field size of $2\rlap.{'}7$.
Maps were made for individual epochs and for the combination of the first, second and fourth epochs. The central frequency (wavelength) in these continuum images is 22.9~GHz (1.31~cm).  The beam sizes and root mean square (rms) noise achieved in the continuum images are given in columns 4--6 of Table \ref{tab:obs}.

For the images of the line data, we first fit and subtract the continuum from the {\it uv} data using the task {\tt uvcontsub}, excluding the inner 900 channels for the fitting. Then, the task {\tt tclean} was used to generate the data cubes of $2\rlap.{'}7$ in size, with 1,000 clean iterations, threshold for cleaning of 25~mJy, and the same pixel size and weighting scheme as the continuum images. The average rms noise in the maps not corrected by the primary beam was 16, 16, 21, and 18~mJy~beam$^{-1}$ in epochs 1, 2, 3, and 4, respectively (Table \ref{tab:obs}). In order to obtain the positions and fluxes of the detected spots at individual channels (c.f.\ Sect. \ref{sec:vla-results}), we perform a 2D gaussian fit to the brightness distribution with the CASA task {\tt imfit}. 

The error in the spot position is given by the astrometric uncertainty, ${\theta_{\rm res}}/ {(2\times{\rm S/N})}$, where $\theta_{\rm res}$ is the FWHM size of the restoring beam,  and S/N the signal-to-noise ratio of the source \citep{Thompson2017}.  The C-configuration maps of the H$_2$O line have an average beam size of $1\rlap.{''}2$. Therefore, for emission detected at S/N=5 the formal (statistical) error in position is $\approx 0\rlap.{''}12$. For the C$\rightarrow$B configuration, the statistical error is $\approx 0\rlap.{''}08$.

\begin{table}
%\tabletypesize{\scriptsize}
\caption{VLBA observed epochs}
\label{tab:obs-vlba} 
%\centering 
{\footnotesize
\begin{tabular}{c c c c c c c}  
\hline\hline  
Epoch & Observation  & Beam  Size             &  P.A.             &  Channel rms            & Observed \\
           & date               & (mas$\times$mas) & ($^{\rm o}$)   & (mJy~beam$^{-1}$)  & Targets    \\
\hline 
%%%%%
A1 & 2020 Mar 20 & 1.4$\times$0.5 & $-13$ & 10 & CARMA-7, 2MASS J18295902--0201575 \\
A2 & 2020 Apr 05  & 1.5$\times$0.4 & $-16$ &  9 & CARMA-7, 2MASS J18295902--0201575 \\
A3 & 2020 Sep  21 & 1.6$\times$0.3 & $-17$ & 9 & CARMA-7 \\
A4 & 2020 Oct  25 & 1.3$\times$0.4 & $-17$ &  8  & CARMA-6 \\
B1 & 2020 Mar 27 & 1.1$\times$0.4 & $-13$ & 10 &  SSTgbs J1830177--021211, SSTgbs J1829053--014156 \\
B2 & 2020 Apr 09 & 1.3$\times$0.6  & $~~~~5$ & 10 & SSTgbs J1830177--021211, SSTgbs J1829053--014156 \\
B3 & 2020 Sep 29 & 1.4$\times$0.4 & $-17$ & 9 &  SSTgbs J1830177--021211, SSTgbs J1829053--014156 \\
B4 & 2020 Nov 01 & 1.4$\times$0.4 & $-17$ & 7 & SSTgbs J1830177--021211 \\ 
\hline 
\end{tabular}
}
\end{table}

%\begin{table*}
%\begin {center}
%\caption{Properties of the VLA detected sources with 22 GHz water emission.}
%\label{tab:line-imfit} 
%%\centering
%\begin{tabular}{c c c c c c D{,}{~\pm~}{-1} D{,}{~\pm~}{4}}
%\hline
%\hline  
% Name&Epoch& $\alpha$ & $\delta$ & $V_{\rm LSR}$&$\Delta V_{\rm LSR}$  & \multicolumn{1}{c}{~~Peak Flux} &\multicolumn{1}{c}{~~~Int.\ Flux}        \\
%           &   & (J2000)  & (J2000)  & (km~s$^{-1}$)    &  (km~s$^{-1}$ )            &  \multicolumn{1}{c}{~(mJy~beam$^{-1}$)}  &  \multicolumn{1}{c}{~~~(mJy)}   \\
%   (1)   &  (2)       &    (3)       &    (4)         &        (5)              &         (6)              &  \multicolumn{1}{c}{~~~~(7)}      &  \multicolumn{1}{c}{~~~(8)}            \\ 
 %
\tabletypesize{\scriptsize}
\begin{deluxetable*}{ccccccD{,}{\pm}{-1}D{,}{\pm}{-1}c}
%\tablenum{1}
\tablecaption{Properties of the VLA detected sources with 22 GHz water emission. \label{tab:line-imfit} }
\tablewidth{0pt}
\tablehead{
\colhead{Name} & \colhead{Epoch} & \colhead{$\alpha$ (J2000)} & \colhead{$\delta$ (J2000)} &  \colhead{$V_{\rm LSR}$}  & 
 \colhead{$\Delta V_{\rm LSR}$} & \colhead{Peak Flux} & \colhead{Int.\ Flux}  & \colhead{$L_{\rm H_2O}$}\\
 \colhead{} & \colhead{} & \colhead{(h:m:s)} & \colhead{($^{\rm o}$:$'$:$''$)} & \colhead{(km~s$^{-1}$)} & \colhead{(km~s$^{-1}$)} & 
 \colhead{(mJy~beam$^{-1}$)} & \colhead{(mJy)} & \colhead{($10^{-10} L_\odot$)}  \\
  \colhead{(1)} & \colhead{(2)} & \colhead{(3)} & \colhead{(4)} & \colhead{(5)} & \colhead{(6)} & 
 \colhead{(7)} & \colhead{(8)} & \colhead{(9)} 
}
%\decimals
\startdata
\hline 
%CARMA-7 \\
% \multirow{3}{*}{CARMA-7} & 1 & 18:30:04.12 & --02:03:02.56 & 10.46 & 1.2 & 96.70,4.30 & 92.83,7.70  \\
%   & 2 & 18:30:04.12 & --02:03:02.56 & 10.46 & 1.4 & 118.27,7.32 & 115.42,12.76  \\
%   & 4 & 18:30:04.12 & --02:03:02.49 & 10.46 & 0.8 & 132.19,6.69 & 172.79,14.34  \\
%   \hline 
%\multirow{4}{*}{SSTgbs J1829053-014156} & 1 & 18:29:05.32 & --01:41:56.93 & -6.18 & 0.8 & 104.32,6.46 & 139.87,14.04  \\
%   & 2 & 18:29:05.33 & --01:41:56.90 & -6.08 & 0.6 & 119.33,6.21 & 89.86,10.17  \\
%   & 3 & 18:29:05.34 & --01:41:56.96 & -6.08 & 0.2 & 126.59,7.72 & 140.77,15.66  \\
%   & 4 & 18:29:05.33 & --01:41:56.99 & -5.97 & 0.6 & 166.97,6.58 & 231.62,14.65  \\
%      \hline 
% \multirow{4}{*}{SSTgbs J1830177-021211} & 1 & 18:30:17.72 & --02:12:11.59 &  6.04 & 0.2 & 78.26,7.64 & 66.09,11.95  \\
%   & 2 & 18:30:17.72 & --02:12:11.64 &  5.30 & 0.6 & 121.87,4.36 & 118.22,7.55  \\
%   & 3 & 18:30:17.72 & --02:12:11.70 &  5.30 & 0.6 & 123.83,6.02 & 126.24,10.61  \\
%   & 4 & 18:30:17.71 & --02:12:11.72 &  5.40 & 0.2 & 86.99,7.78 & 71.35,12.95  \\
 % 
 \multirow{3}{*}{CARMA-7$^a$}  &  1  &  18:30:04.12  &  --02:03:02.56  &  10.46  &  1.2  &  96.70,4.30  &  92.83,7.70  &  5.6  \\
  &  2  &  18:30:04.12  &  --02:03:02.56  &  10.46  &  1.4  &  118.27,7.32  &  115.42,12.76  &  6.0  \\
  &  4  &  18:30:04.12  &  --02:03:02.49  &  10.46  &  0.8  &  132.19,6.69  &  172.79,14.34  &  5.7  \\
  \hline
\multirow{4}{*}{SSTgbs J1829053-014156}  &  1  &  18:29:05.32  &  --01:41:56.93  &  -6.18  &  0.8  &  104.32,6.46  &  139.87,14.04  &  4.1  \\
  &  2  &  18:29:05.33  &  --01:41:56.90  &  -6.08  &  0.6  &  119.33,6.21  &  89.86,10.17  &  2.9  \\
  &  3  &  18:29:05.34  &  --01:41:56.96  &  -6.08  &  0.2  &  126.59,7.72  &  140.77,15.66  &  1.6  \\
  &  4  &  18:29:05.33  &  --01:41:56.99  &  -5.97  &  0.6  &  166.97,6.58  &  231.62,14.65  &  4.7  \\
  \hline
\multirow{4}{*}{SSTgbs J1830177-021211}  &  1  &  18:30:17.72  &  --02:12:11.59  &  6.04  &  0.2  &  78.26,7.64  &  66.09,11.95  &  0.6  \\
  &  2  &  18:30:17.72  &  --02:12:11.64  &  5.30  &  0.6  &  121.87,4.36  &  118.22,7.55  &  2.2  \\
  &  3  &  18:30:17.72  &  --02:12:11.70  &  5.30  &  0.6  &  123.83,6.02  &  126.24,10.61  &  2.4  \\
  &  4  &  18:30:17.71  &  --02:12:11.72  &  5.40  &  0.2  &  86.99,7.78  &  71.35,12.95  &  1.2  \\
  \hline 
\enddata
\tablecomments{Column 1 is the source name. Column 2 is the observed epoch; Column 3 and 4 give the coordinates of the weighted mean position of all contributing emission spots; Columns 5 and 6 are the velocity of the highest intensity channel, and velocity range of the H$_2$O emission; Column 7 and 8 give the peak and integrated flux density, respectively, at the highest intensity channel and associated errors obtained using {\tt imfit}. Column 9 gives the water maser luminosity. }
\tablenotetext{a}{Due to failures during the observations, CARMA-7 was not observed in the third epoch. }
\end{deluxetable*}
%
%\end{tabular}
%\tablefoot{Column 1 is the source name. Column 2 is the observed epoch; Column 3 and 4 give the coordinates of the weighted mean position of all contributing emission spots; Columns 5 and 6 are the velocity of the highest intensity channel, and velocity range of the H$_2$O emission; Column 7 and 8 give the peak and integrated flux density, respectively, at the highest intensity channel and associated errors obtained using {\tt IMFIT}.
%%\tablefoottext{a}{The quoted errors are from {\tt IMFIT}. } 
%%
%}
%\end{center}
%\end{table*}

\subsection{VLBA observations}\label{sec:obs-vlba}

We conducted multi-epoch Very Long Baseline Array (VLBA) observations toward 4 targets, including the 3 sources that were undoubtedly detected in H$_2$O emission with the VLA (Sect. \ref{sec:vla-results}), and one more star with tentative detection (2MASS J18295902--0201575). These observations were conducted between March and November, 2020 as part of program BO061 (Table \ref{tab:obs-vlba}). The data were taken at 22.2 GHz with 4 intermediate frequency (IF) bands, each of 16~MHz bandwidth. One IF was centered at the $6_{1,6} - 5_{2,3}$ H$_2$O transition and correlated with a channel spacing of $\sim$0.2~km~s$^{-1}$ (15.625~kHz). We observed the quasar J1824+0119 ($\alpha$(J2000)=18:24:48.143436, $\delta$(J2000)=+01:19:34.20183) as the phase reference calibrator, which we alternated with target observations in fast switching mode, switching sources every $\approx$30 seconds. Additional 30-min blocks of calibrators distributed over a wide range of elevations were observed at 23.7~GHz every $\approx$2 hours during each $\approx$9-hr observing run. The observations were organized in two blocks, ``A'' and ``B'', with each block observing up to 2 targets.  The total on-source time of the water masers was 1.5 and 3~hr for blocks observing two and one targets, respectively (see Table \ref{tab:obs-vlba}). 
The blocks have been observed in a total of 4 epochs as of the year 2020 (labeled as 1 to 4 in Table \ref{tab:obs-vlba}). Here, we only report the detections achieved so far (Sect. \ref{sec:masers-vlba}). The full analysis of the multi-epoch VLBA data will be presented in a forthcoming paper. 

Data calibration was performed with the Astronomical Imaging System (AIPS; \citealt{Greisen2003}), using the ParselTongue scripting interface \citep{Kettenis2006} and following standard procedures for phase-referencing observations \citep[e.g.,][]{Reid2009}. Since the VLA-detected masers are relatively weak (<1~Jy; Section \ref{sec:detections}), the fringe-fitting solutions were derived from the scans on the phase-reference calibrator and then applied to the target. Once the calibration was completed, we imaged individual channel maps of $4096\times4096$ pixels using a pixel size of $50~\mu$as. Spot positions and fluxes were determined by fitting a Gaussian to the brightness distribution at individual channels using the AIPS task {\tt jmfit}. The  expected statistical positional errors are of the order of 70~$\mu$as for emission detected at S/N=5.

\section{Results}\label{sec:detections}

\begin{figure*}[!bht]
\begin{center}
 {\includegraphics[width=0.32\textwidth,angle=0]{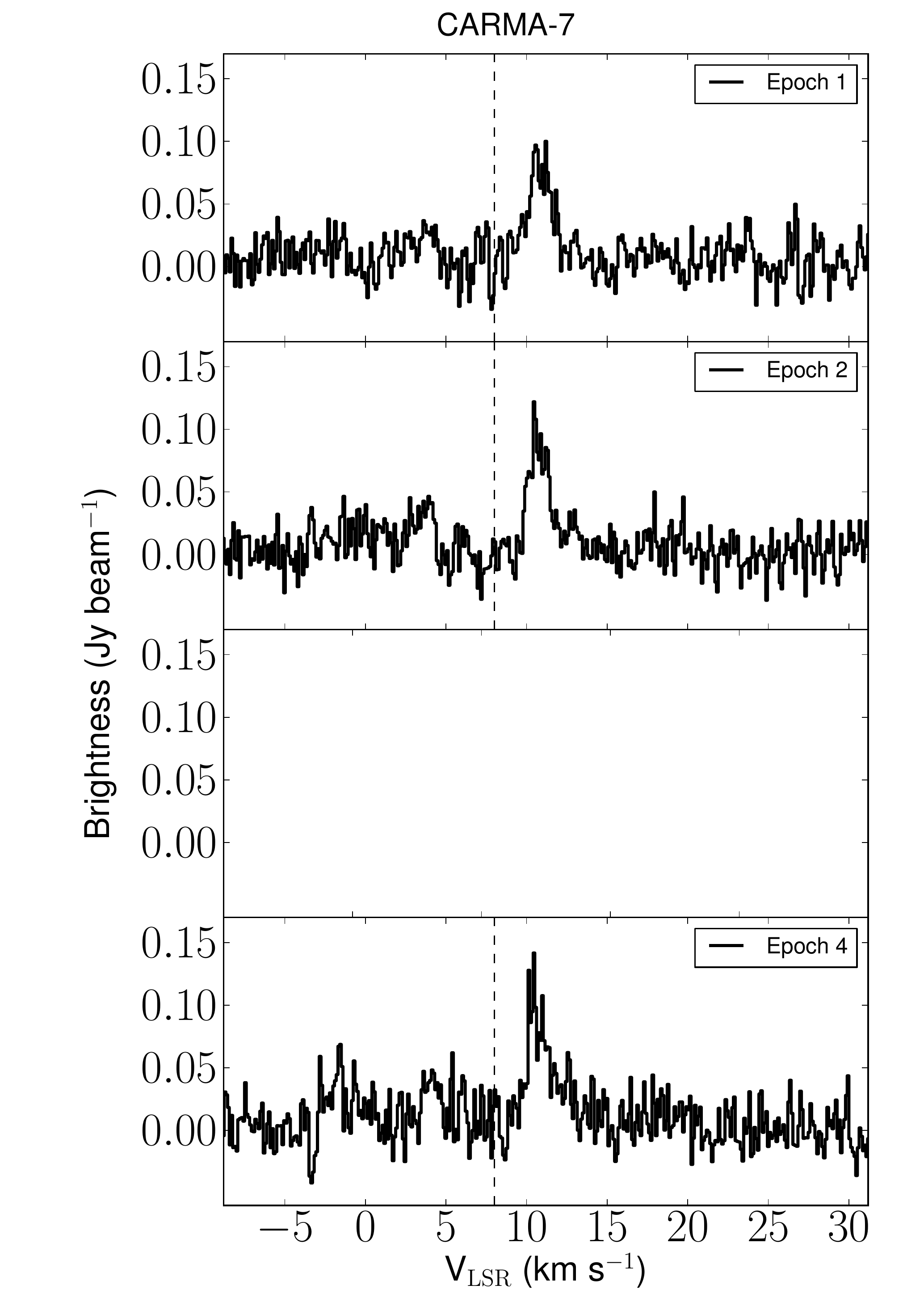}}
 {\includegraphics[width=0.32\textwidth,angle=0]{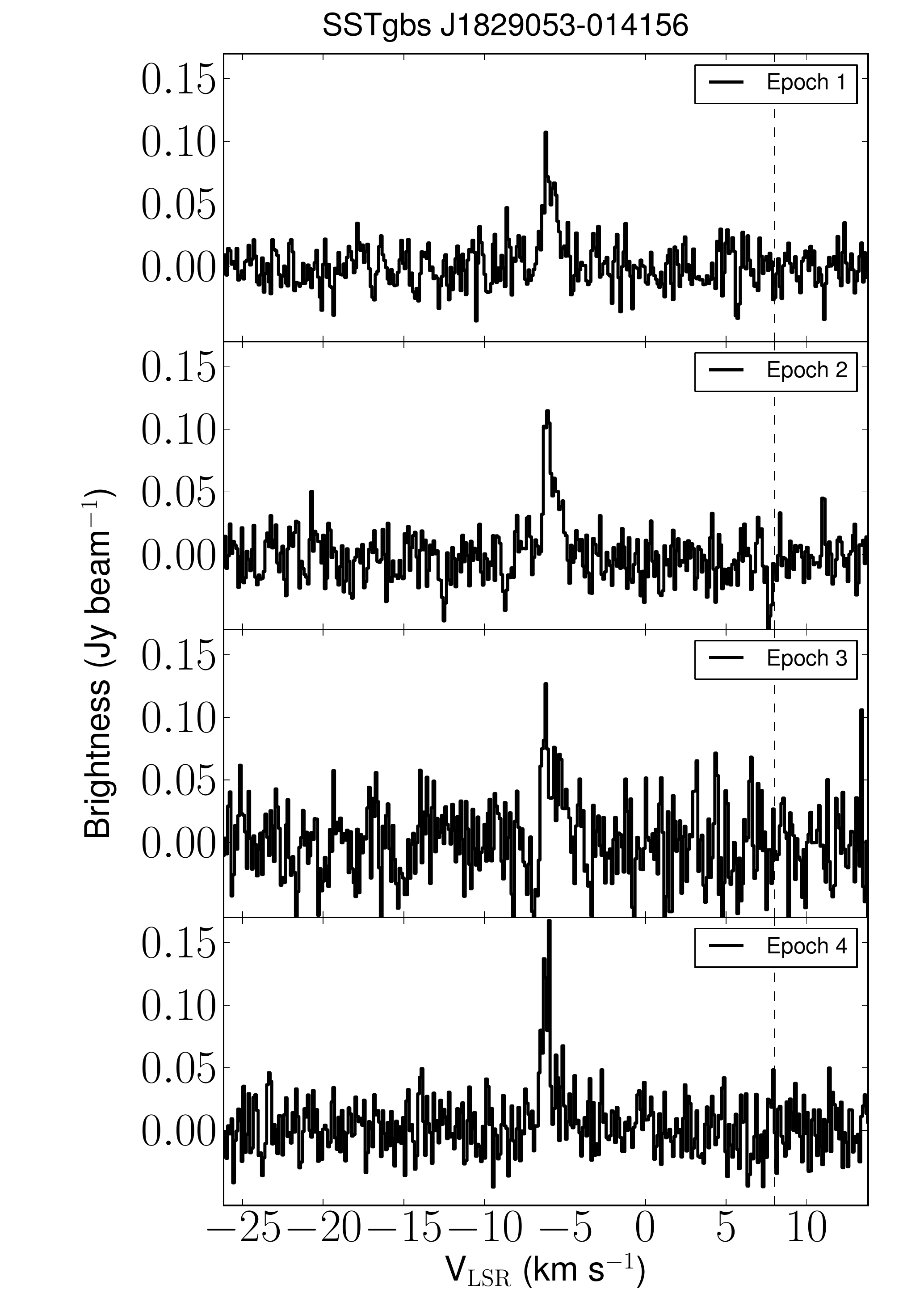}}
 {\includegraphics[width=0.32\textwidth,angle=0]{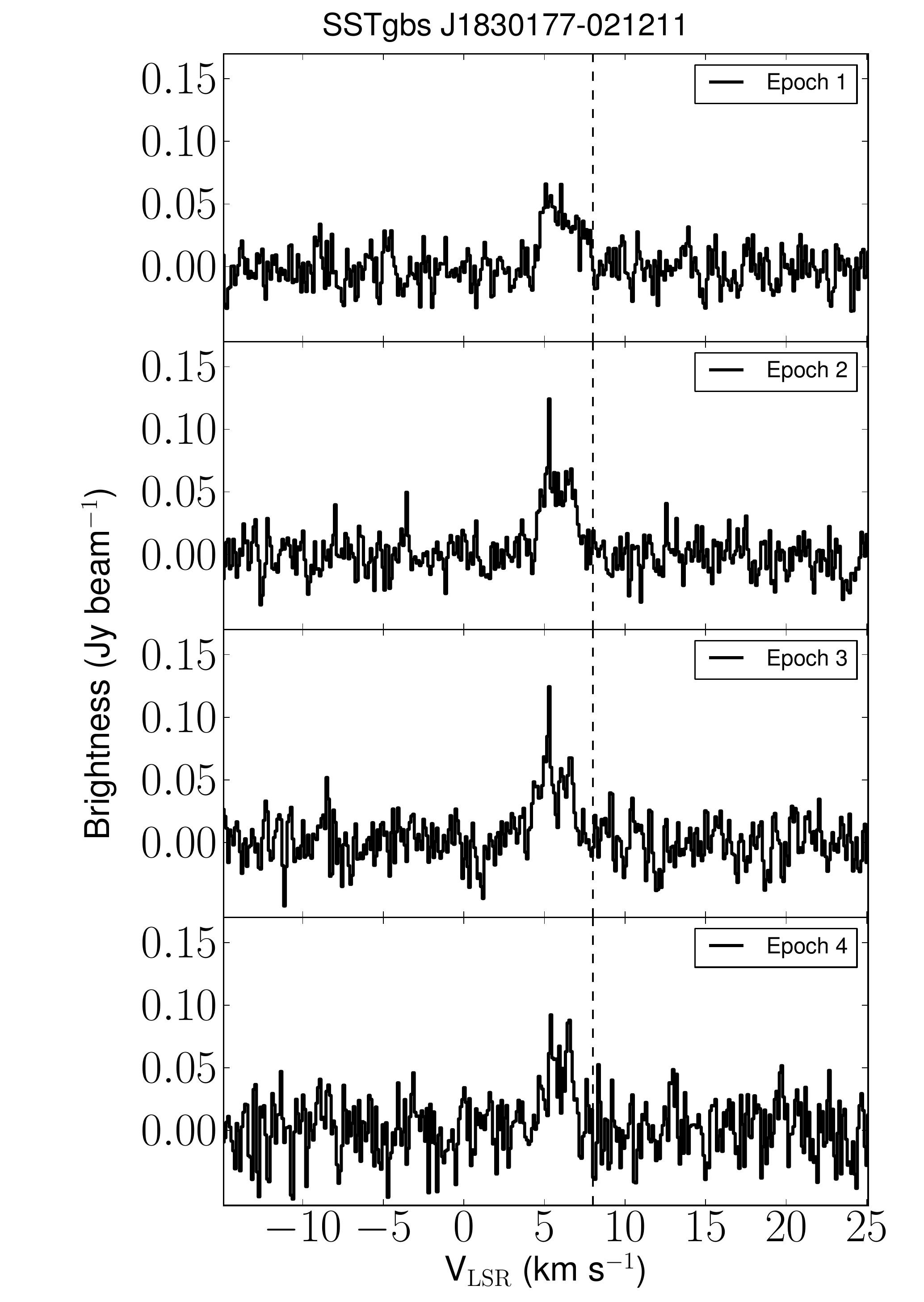}}
 \end{center}
\caption{Spectra of H$_2$O emission detected with the VLA extracted at the peak pixel from the data cubes. The epoch of observation is indicated in the legends. 
The vertical dashed line at  $V_{\rm LSR} = 8$~km~s$^{-1}$ marks the systemic velocity of the cloud. 
Due to failures during the observations, several fields, including CARMA-7, were skipped in the third epoch.
}
\label{fig:vla-spectra}
\end{figure*}

\subsection{VLA detected sources with H$_2$O emission}\label{sec:vla-results}

The cubes of the H$_2$O line were visually inspected to search for emission at the location of the targets.
Only 3 sources have detected H$_2$O emission, whose properties are listed in Table \ref{tab:line-imfit}. Column 1 of this table indicates
the name of the source. Column 2 gives the epoch of detection. Column 3 and 4 give the mean position obtained by taking the weighted mean of the contributing maser spots, where ``spot'' refers to emission detected in a single channel map. Column 5 gives line-of-sight LSR velocity of the channel with the highest intensity.  Column 6 gives the velocity range of H$_2$O emission. Columns 7 and 8 indicate peak and integrated flux density of the highest-intensity channel, respectively.  Column 9 gives the water maser luminosity.  

In the three detected sources, the H$_2$O emission is weak, with fluxes below $\sim$230~mJy and velocity spread $\lesssim$1~km~s$^{-1}$. 
Figure~\ref{fig:vla-spectra} presents extracted spectra at the position with highest intensity.
From this figure, it is clear that the emission shows variability in both flux and velocity (the velocity range of H$_2$O emission changes between epochs). This, together with the narrow widths of the lines, which are in the range from $0.7$ to $2.5$~km~s$^{-1}$, suggest that the detected H$_2$O emission is presumably due to masers. 
We also notice that  spatially distinct groups of spots contribute to the observed H$_2$O spectra toward SSTgbs J1829053--014156 and SSTgbs J1830177--021211. These groups of spots may correspond to spatially separated features, where ``feature'' refers to emission observed in contiguous velocity channels at nearly the same position. Since the poor angular resolution of the VLA does not allow us to  unambiguously separate the features, we average all maser spots for the positions reported in Table \ref{tab:line-imfit}, ignoring the possibility that they may be part of distinct features. \\

In the following, we discuss each detected source separately.  \\

\begin{figure}[!bht]
\begin{center}
{\includegraphics[width=0.45\textwidth,angle=0]{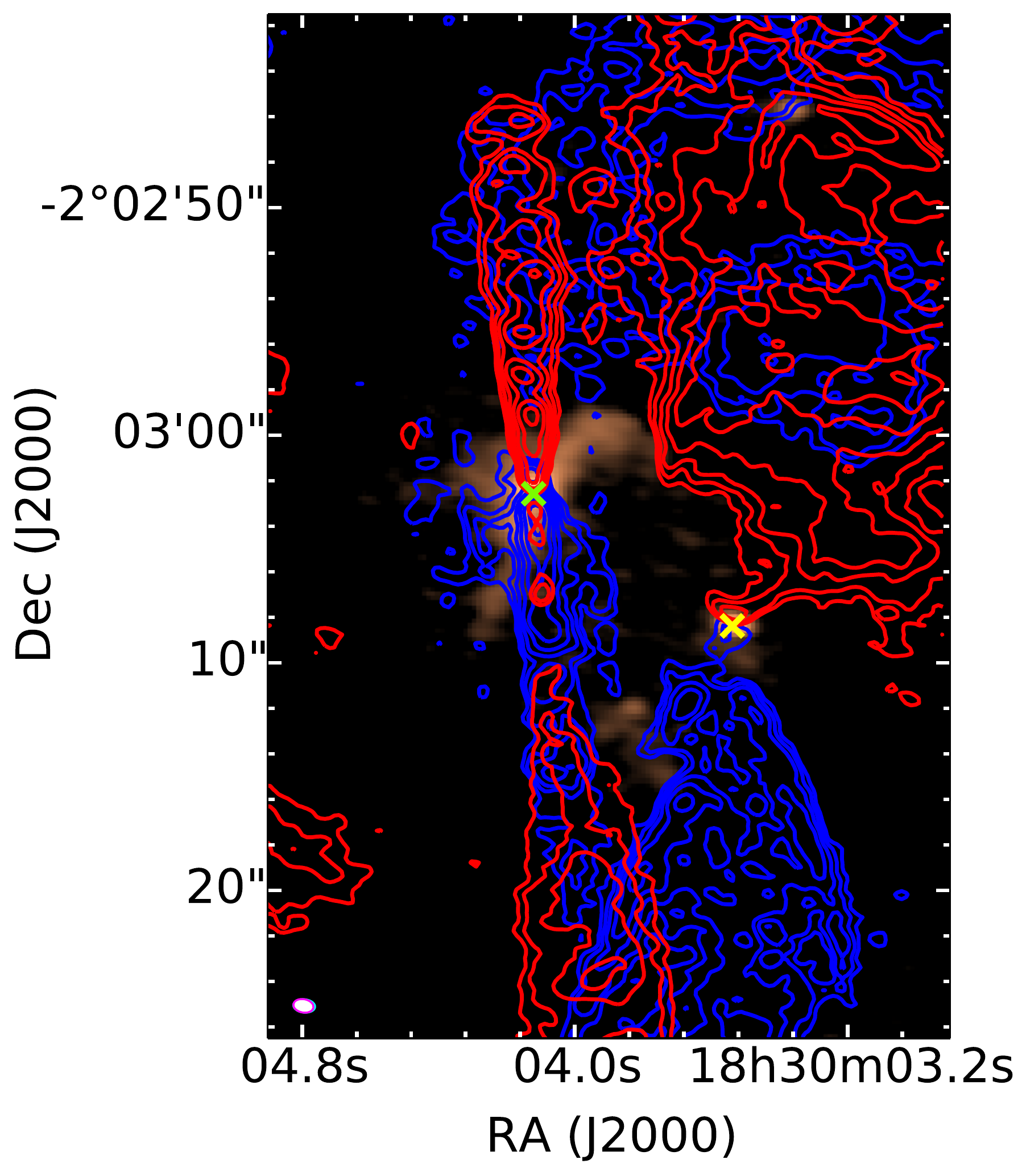}} 
{\includegraphics[width=0.45\textwidth,angle=0]{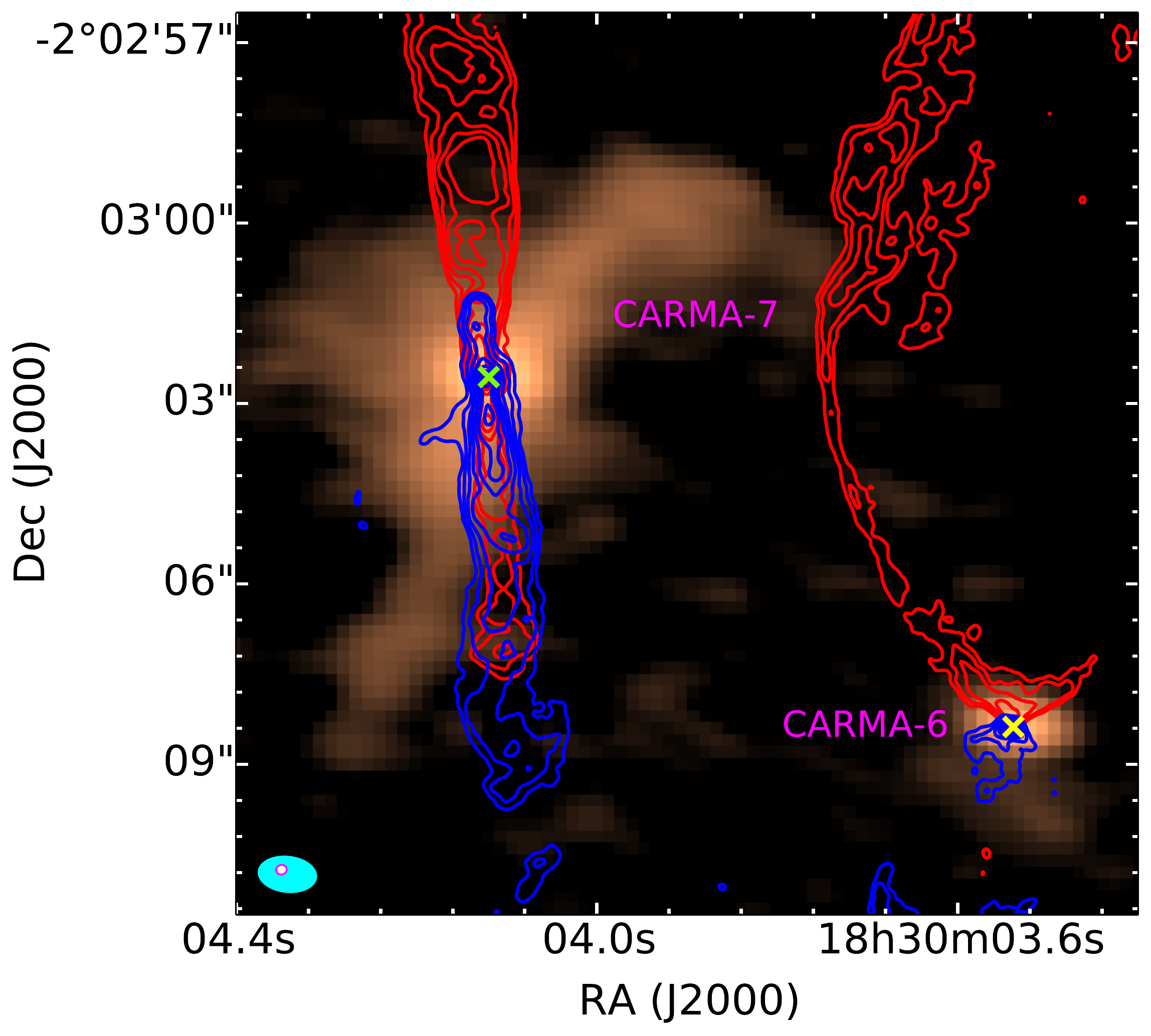}}
 \end{center}
\caption{Large scale molecular outflows traced by CO ($J = 2-1$) at 230.538 GHz from ALMA observations toward CARMA-7 and CARMA-6 \citep{Plunkett_2015}. The integration ranges are  $-$20 to 4~km~s$^{-1}$  for the blue-shifted component and 12 to 40~km~s$^{-1}$ for red-shifted component. The $n$th contour is at $\left(\sqrt{2}\right)^{n}\times S_{\rm max} \times p$, where $S_{\rm max}=3.5$ and 6.3~Jy~beam$^{-1}$~km~s$^{-1}$ (for blue-shifted and red-shifted emission, respectively), $n$=0, 1, 2, 3, 4 ..., and $p$=10\%. The background is an ALMA map of 1-mm continuum emission \citep{Plunkett2018}. 
The right panel shows a zoom-in of the central part of the mapped region. The contours correspond to CO ($J = 3-2$) emission at 345.796 GHz from ALMA observations, integrated in the same velocity range as the CO ($J = 2-1$) data, with $S_{\rm max}=10.8$~Jy~beam$^{-1}$~km~s$^{-1}$. 
In both panels, the ``X''s indicate the average position of the water masers detected with the VLA (green)
and VLBA (yellow; see Sect. \ref{sec:masers-vlba}).
The beamsizes are shown in the bottom left corner of the images as white (molecular data) and cyan (continuum emission) ellipses.
}
\label{fig:outflows}
\end{figure}

{\bf CARMA-7}. Also known as SerpS-MM18a \citep{Maury2019}, it is a Class 0 protostar \citep{Maury2011} with strong millimeter continuum emission \citep{Plunkett2015ApJ} and a highly collimated bipolar outflow extending $\sim$0.16~pc \citep{Plunkett_2015}. Several  knots are seen  along the outflow, suggesting episodic events that are attributed to variations in the accretion rate of mass onto the protostar.  There is a nearby protostar, CARMA-6 (also known as SerpS-MM18b; \citealt{Maury2019}), located to the southwest of CARMA-7, which also has millimeter continuum emission and is classified as a Class 0+I object \citep{Kern2016AJ}. The molecular outflow associated with CARMA-6 has a much wider opening angle (see Fig. \ref{fig:outflows}). The dust masses of CARMA-7 and CARMA-6 are estimated to be 1.21 and 0.43~$M_\odot$, respectively \citep{Plunkett2015ApJ}. They have {\it internal} luminosities of 13 and 16~$L_\odot$ that were derived from the 70-$\mu$m band of {\it Herschel} assuming a distance of 350~pc \citep{Podio2020}.  These luminosities are rescaled to 20 and 25~$L_\odot$ for a distance of 436~pc.

The water maser detected with the VLA toward  CARMA-7 is found at the very base of the CO ($J$=2--1) molecular outflow (see Fig. \ref{fig:outflows}) traced by ALMA \citep{Plunkett_2015}. This position also coincides with the peak of the millimeter continuum emission (see right panel of Fig. \ref{fig:outflows}). The velocity-integrated intensity map of the CO ($J$=3--2)  line is also shown in this figure (right panel). In CARMA-6, the red-shifted CO ($J$=3--2) outflow seems to correspond to the cavity walls of the CO ($J$=2--1)  outflow.
Radio continuum sources associated with both CARMA-7 and CARMA-6 were found by  \cite{Kern2016AJ} from observations at 4.75--7.25~GHz (their sources VLA~12 and VLA~13). The radio continuum emission is also detected in our observations (see Sect. \ref{sec:cont} and Fig. \ref{fig:vla-continuum-all} in Appendix \ref{sec:appendix};  sources \#10 and \#9). \cite{Kern2016AJ} derived radio spectral indices of $2.31\pm0.12$ and $0.51\pm0.08$ for CARMA-7 and CARMA-6, respectively, which are indicative of thermal radio emission from ionized gas, and proposed that the radio emission is tracing the base of collimated outflows.  \\ 
%% VLA~12 = CARMA-7 = 10
%% VLA~13 = CARMA-6 = 9

{\bf SSTgbs J1829053--014156/IRAS~18264-0143}.  This object is also a known YSO \citep{Dunham_2015}. The extinction corrected slope of the infrared spectral energy distribution (SED) is 0.96,
which places the source in the Class 0+I phase \citep{Dunham_2015}.  The stellar extinction corrected bolometric luminosity is $L_{\rm Bol} = 2.9~L_\odot$ obtained by assuming a distance of 260~pc \citep{Dunham_2015}. This value is rescaled to $8.2~L_\odot$ for a distance of 436~pc. There is a 1.2 mm continuum peak close (at $\approx6''$) to the water maser, called Aqu-MM3, which was identified as a Class 0+I object \citep{Maury2011}.  A dust mass of 1.7~$M_\odot$ and a bolometric luminosity of 14.3~$L_\odot$ (corrected for the assumed distance) was measured for the millimeter continuum source. 
We detected radio continuum emission associated to this source (see Fig. \ref{fig:vla-continuum-all} in Appendix \ref{sec:appendix}, source \#2), which may be tracing the base of the jet. The maser position coincides, within the position errors, with the peak of radio continuum (Fig. \ref{fig:vla-continuum-all}).

\begin{figure*}[!bht]
\begin{center}
 {\includegraphics[width=0.7\textwidth,angle=0]{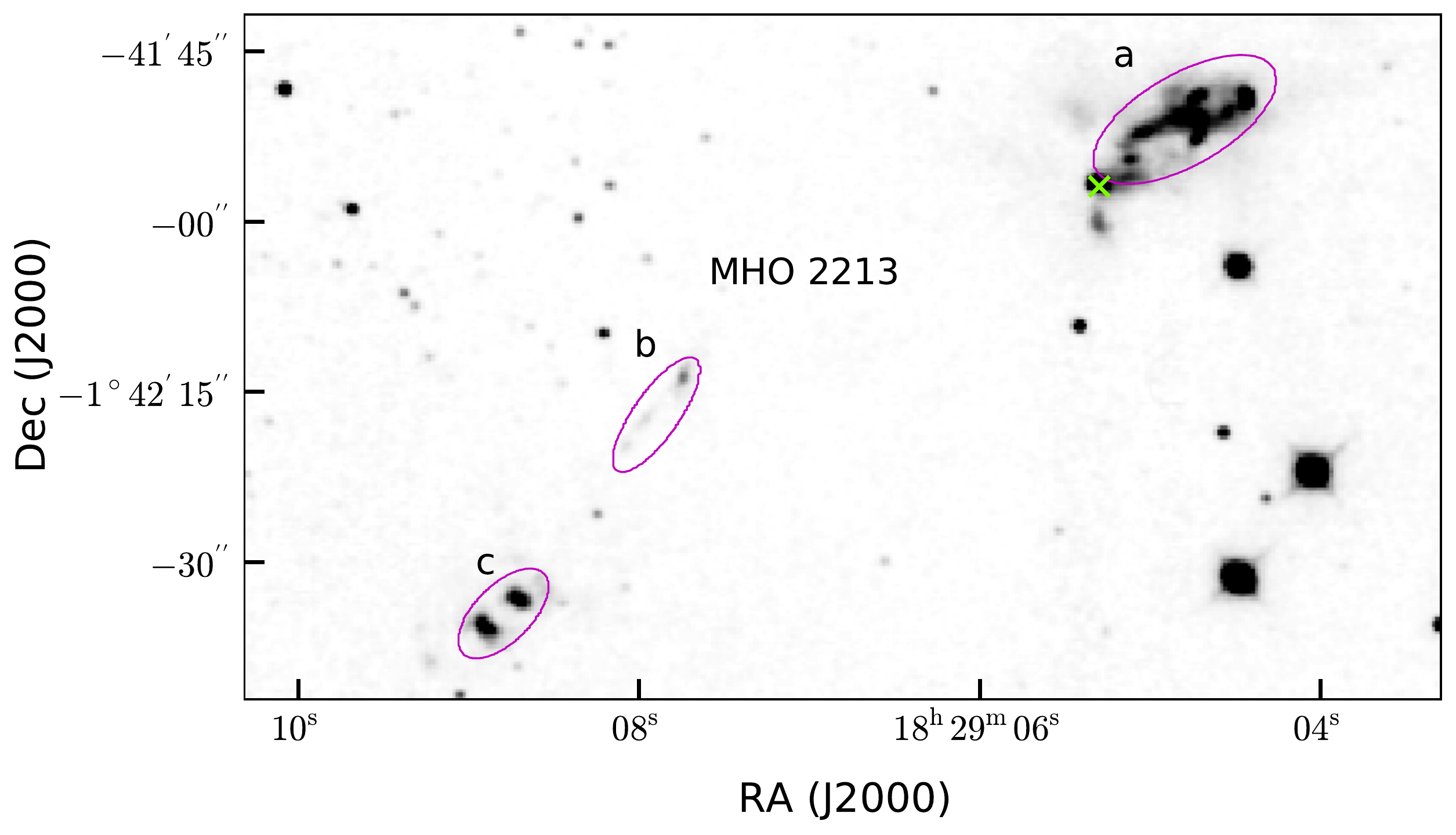}}
 \end{center}
\caption{H$_2$ 2.12 $\mu$m image of MHO2213, the outflow associated with SSTgbs J1829053--014156/IRAS 18264--0143 \citep{Zhang2015}. The MHO features are marked with magenta ellipses and denoted with letters. 
The green ``X'' denotes the position of the maser, which coincides with the position of the outflow driving source. }
\label{fig:outflow-J1829053}
\end{figure*}

 We searched the literature for molecular outflows that can be associated with this maser source. Observations of the CO ($J=3-2$) transition at 345.796~GHz were conducted by \cite{Nakamura2011}  with the ASTE 10~m telescope to study the outflow activity in Serpens South. In their images, there is a clear bipolar CO outflow in the vicinity of SSTgbs J1829053--014156 (the red-shifted and blue-shifted outflow components are called R6 and B11, in the nomenclature of \citealt{Nakamura2011}). The maser is very close to the base of the blue-shifted component (see Fig.\ 9 of \citealt{Nakamura2011}).  This outflow is also traced by H$_2$ emission at 2.12~$\mu$m \citep{Davis2010,Zhang2015}. The associated molecular hydrogen emission-line object is MHO~2213, which is thought to be driven by IRAS~18264--0143. The position angle of $118^{\rm o}$ of the MHO is similar to the orientation of the CO outflow (see Fig. \ref{fig:outflow-J1829053}). The maser is located at the base of the MHO feature that is associated with the blue-shifted CO lobe.  \\

{\bf SSTgbs J1830177--021211/IRAS~18276--0214}. This object is a known YSO \citep{Dunham_2015,Winston2018AJ}. The infrared SED has a extinction corrected slope of $-2.22$ \citep{Dunham_2015},   
which places it in the Class III phase. Later, based on its infrared colors, \cite{Winston2018AJ} classified it as a 
disk-bearing pre-main-sequence object (equivalent to the Class II/transition disk class of \citealt{Dunham_2015}).
The extinction corrected bolometric luminosity is $L_{\rm Bol} = 158~L_\odot$, which has been rescaled for a distance of 436~pc. Its mass has not been estimated.

 \begin{figure}[!bht]
 \begin{minipage}{\textwidth}
\begin{center}
{\includegraphics[width=0.48\textwidth,angle=0]{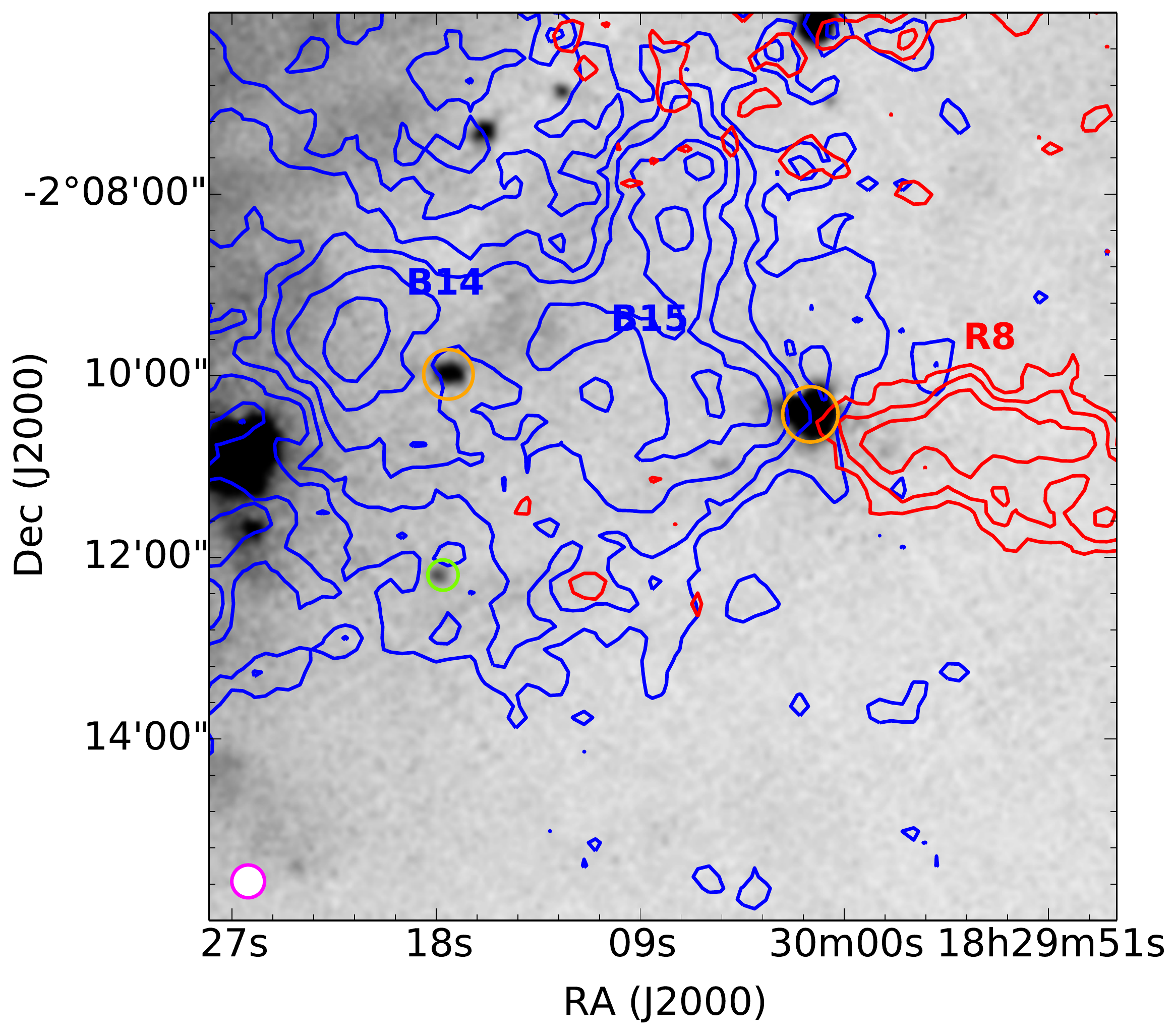}}
 \end{center}
\caption{Molecular outflow lobes traced by CO ($J=1-0$) at 115.27~GHz \citep{Nakamura2019} toward SSTgbs J1830177--021211. The integration ranges are  $-$15 to 4~km~s$^{-1}$  for the blue-shifted component and 11 to 30~km~s$^{-1}$ for red-shifted component. The labels denote lobes identified by \cite{Nakamura2011} from CO ($J=3-2$) observations at 345.796~GHz.  The $n$th contour is at $\left(\sqrt{2}\right)^{n}\times S_{\rm max} \times p$, where $S_{\rm max}=0.2$~K~km~s$^{-1}$, $n$=0, 1, 2, 3, 4 ..., and $p$=11\%. In the background we show a {\it Herschel} 70 $\mu$m map retrieved from the Science Archive ({\url{http://archives.esac.esa.int/hsa/whsa/}}). 
Orange circles mark the location of the {\it Herschel} protostellar cores \citep{Konyves2015} that have been identified as the outflow driving sources \citep{Nakamura2019}. The source with detected H$_2$O masers is indicated by the green circle. 
The beamsize is shown in white in the bottom left corner.
}
\label{fig:outflows2}
\end{minipage}
\end{figure}

\begin{deluxetable*}{cccccD{,}{\pm}{-1}D{,}{\pm}{-1}cc}
%\tablenum{1}
\tablecaption{Properties of the maser features detected with the VLBA. \label{tab:astro-vlba} }
\tablewidth{0pt}
\tablehead{
\colhead{Source} & \colhead{$\alpha(J2000)^{a}$} & \colhead{$\delta(J2000)^{a}$} & \colhead{Feature} &
 \colhead{Epoch} & \colhead{R.A.\ offset} & \colhead{Dec. offset} & \colhead{$V_{\rm LSR}$$^{b}$} & \colhead{$L_{\rm H_2O}$} \\
 \colhead{} & \colhead{(h:m:s)} & \colhead{($^{\rm o}$:$'$:$''$)} & \colhead{\# } &
 \colhead{} & \colhead{(mas)} & \colhead{(mas)} & \colhead{(km~s$^{-1}$)} & \colhead{($10^{-10}~L_\sun$)} \\
   \colhead{(1)} & \colhead{(2)} & \colhead{(3)} & \colhead{(4)} & \colhead{(5)} & \colhead{(6)} & 
 \colhead{(7)} & \colhead{(8)} & \colhead{(9)} 
}
%\decimalcolnumbers
\startdata
\multirow{10}{*}{ SSTgbs J1830177--021211}&\multirow{3}{*}{ 18:30:17.7141494}&\multirow{3}{*}{ --02:12:11.685739}&1&B1&-0.00,0.00&0.00,0.06&6.2&\multirow{3}{*}{ 46}\\
&&&2&B1&0.72,0.02&0.98,0.04&4.6\\
&&&3&B1&1.48,0.04&0.68,0.09&5.3\\ \cline{2-9}
&\multirow{4}{*}{ 18:30:17.7141433}&\multirow{4}{*}{ --02:12:11.685734}&1&B2&0.00,0.04&0.00,0.03&6.2&\multirow{4}{*}{218}\\
&&&2&B2&0.79,0.01&0.89,0.01&4.6\\
&&&3&B2&1.31,0.11&0.81,0.04&5.5\\
&&&4&B2&1.75,0.07&1.07,0.03&6.6\\ \cline{2-9}
&18:30:17.7142092&--02:12:11.685789&1&B3&0.00,0.05&0.00,0.06&6.2&  5\\ \cline{2-9}
&18:30:17.7142152&--02:12:11.686065&1&B4&0.00,0.05&0.00,0.04&6.3& 17\\
\hline
% \colhead{} & \colhead{$\alpha(J2000)^{b}$} & \colhead{$\delta(J2000)^{b}$} & \colhead{} &
% \colhead{} & \colhead{} & \colhead{} & & \colhead{$L_{\rm H_2O}$} \\
%  \colhead{} & \colhead{(h:m:s)} & \colhead{($^{\rm o}$:$'$:$''$)} & \colhead{} &
% \colhead{} & \colhead{} & \colhead{} & & \colhead{($10^{-6}~L_\sun$)} \\
%  \hline
\multirow{4}{*}{ CARMA--6}&\multirow{4}{*}{ 18:30:03.5380}&\multirow{4}{*}{ --02:03:08.377}&1&A4&-3.47,0.06&2.85,0.06&11.3&\multirow{4}{*}{1.2}\\
&&&2&A4&-2.81,0.05&2.43,0.07&10.2\\
&&&3&A4&-0.94,0.01&0.06,0.15&9.4\\
&&&4&A4&-0.00,0.01&0.01,0.18&8.5\\
\enddata
%\tablecomments{}
\tablenotetext{a}{Reference position at the given epoch.}
\tablenotetext{b}{Line-of-sight velocity of the feature obtained as the intensity-weighted mean $V_{\rm LSR}$ of the contributing spots. }
\end{deluxetable*}

The detection of water maser emission in this source is unexpected given that earlier surveys have suggested that maser activity disappears after the main accretion and outflow (Class 0--Class I) phase \citep{Furuya2001}. The maser has also been detected in our follow-up VLBA observations (see Sect. \ref{sec:masers-vlba}). We did not detect radio continuum emission associated to the maser and no radio continuum has been reported in the literature either. There are three molecular outflow lobes (B14, B15 and R8 in the nomenclature of \citealt{Nakamura2011}) in the surroundings of the water maser as seen in Fig. \ref{fig:outflows2}, where we show CO ($J=1-0$) data at 115.27~GHz taken with the Nobeyama telescope \citep{Nakamura2019}. The outflow lobes identified by \cite{Nakamura2011} are indicated in this figure, as well as the positions of the putative driving sources, which are taken from the {\it Herschel} catalog of protostellar cores \citep{Konyves2015}. The CO ($J=1-0$) emission at the position of the maser is relatively weak. The H$_2$ 2.12~$\mu$m image is dominated by very strong emission from IRAS~18276--0214 (Fig.  12.9 in \citealt{Zhang2015}), so it is difficult to find an association with an H$_2$ outflow feature. \\

\subsection{VLBA detected sources with H$_2$O maser emission}\label{sec:masers-vlba}
 
\begin{figure*}[!bht]
\begin{center}
 {\includegraphics[width=0.458\textwidth,angle=0]{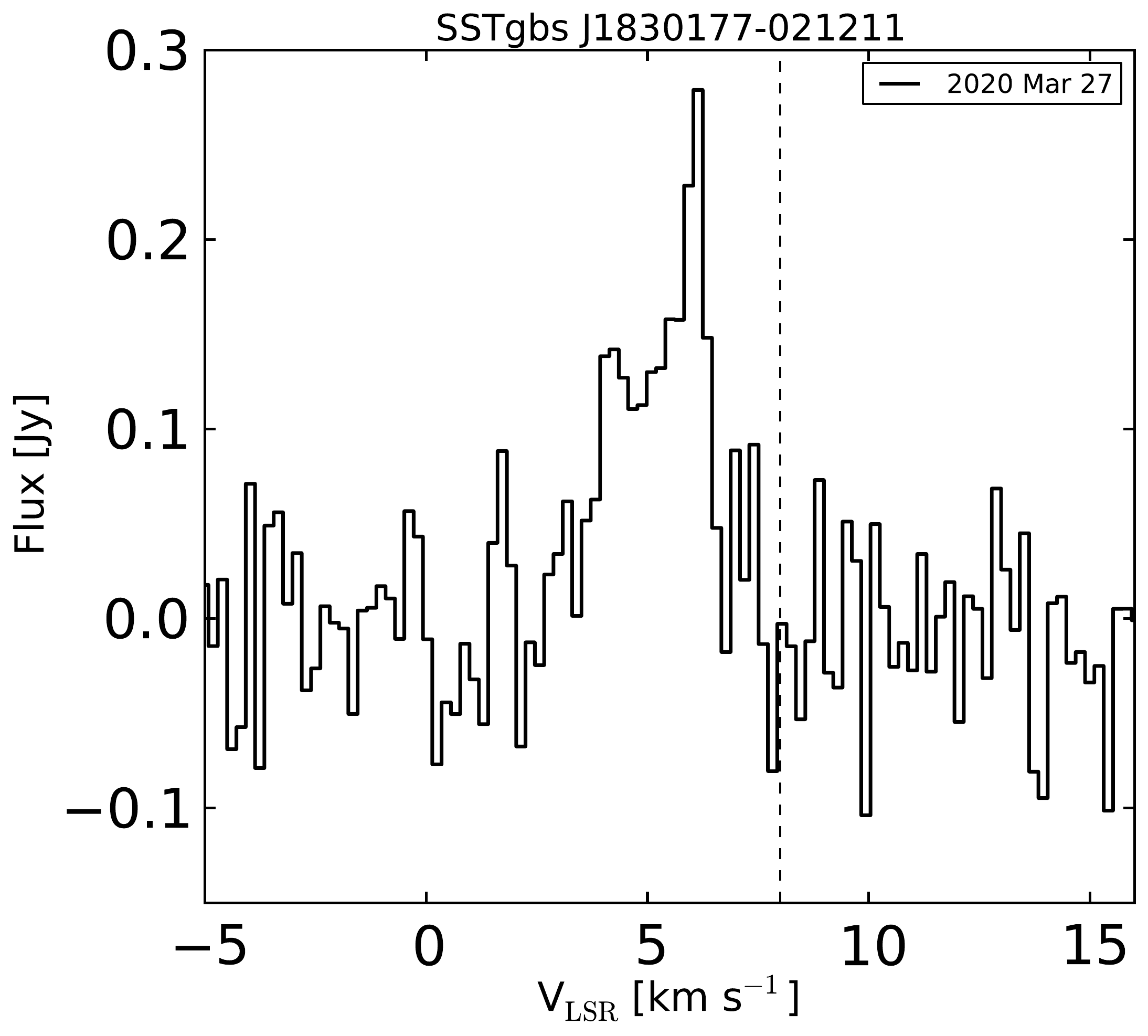}}
 {\includegraphics[width=0.45\textwidth,angle=0]{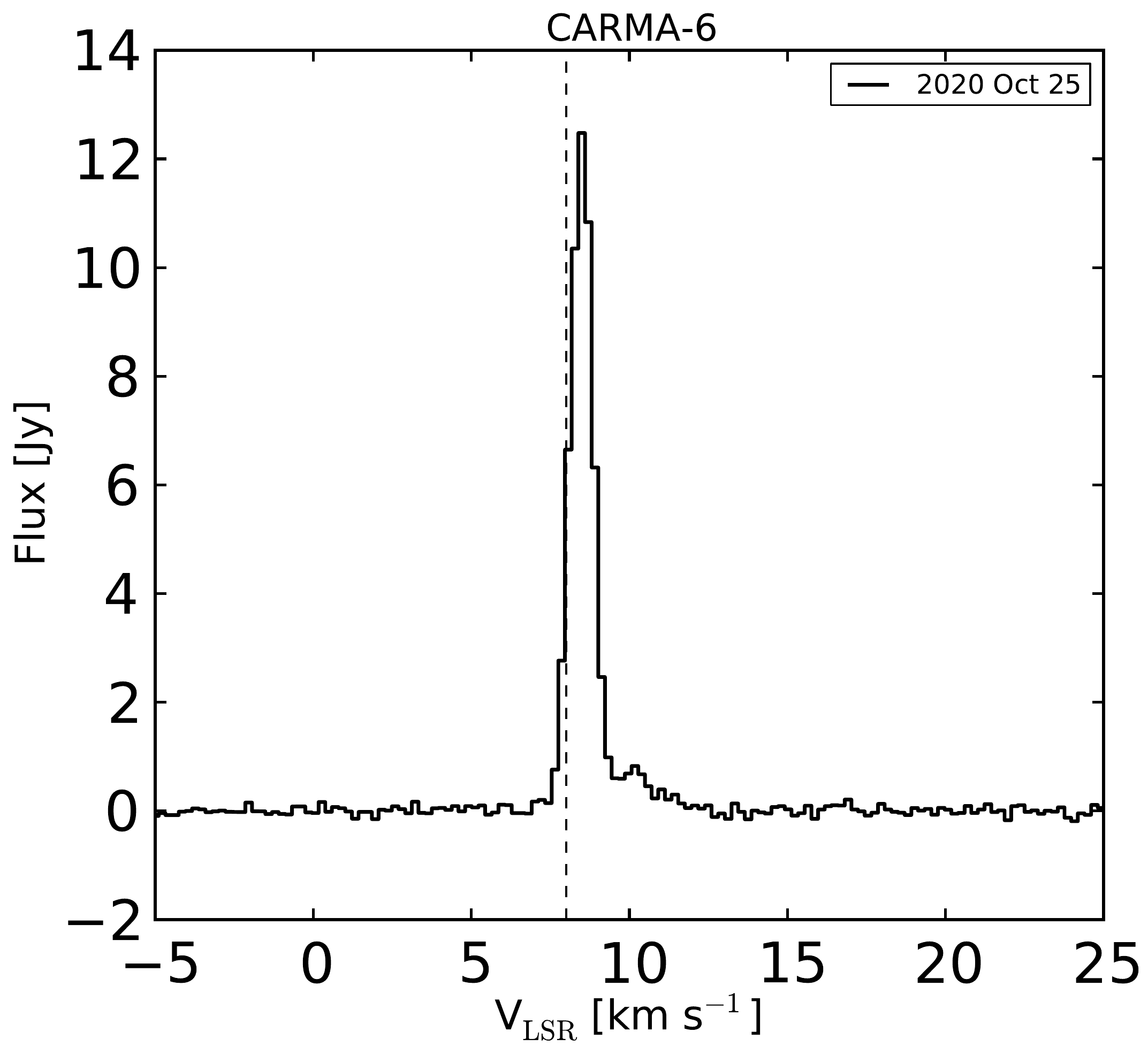}}
 \end{center}
\caption{VLBA spectra of the 22 GHz H$_2$O maser transition toward  SSTgbs~J1830177--021211/IRAS 18276--0214 (left) and CARMA--6 (right) obtained by integrating over an area that covers all detected spots. The legends indicate the epoch of observation. The vertical dashed line at  $V_{\rm LSR} = 8$~km~s$^{-1}$ marks the systemic velocity of the cloud. }
\label{fig:vlba-spectra}
\end{figure*}

\begin{figure*}[!bht]
\begin{center}
 {\includegraphics[width=0.45\textwidth,angle=0]{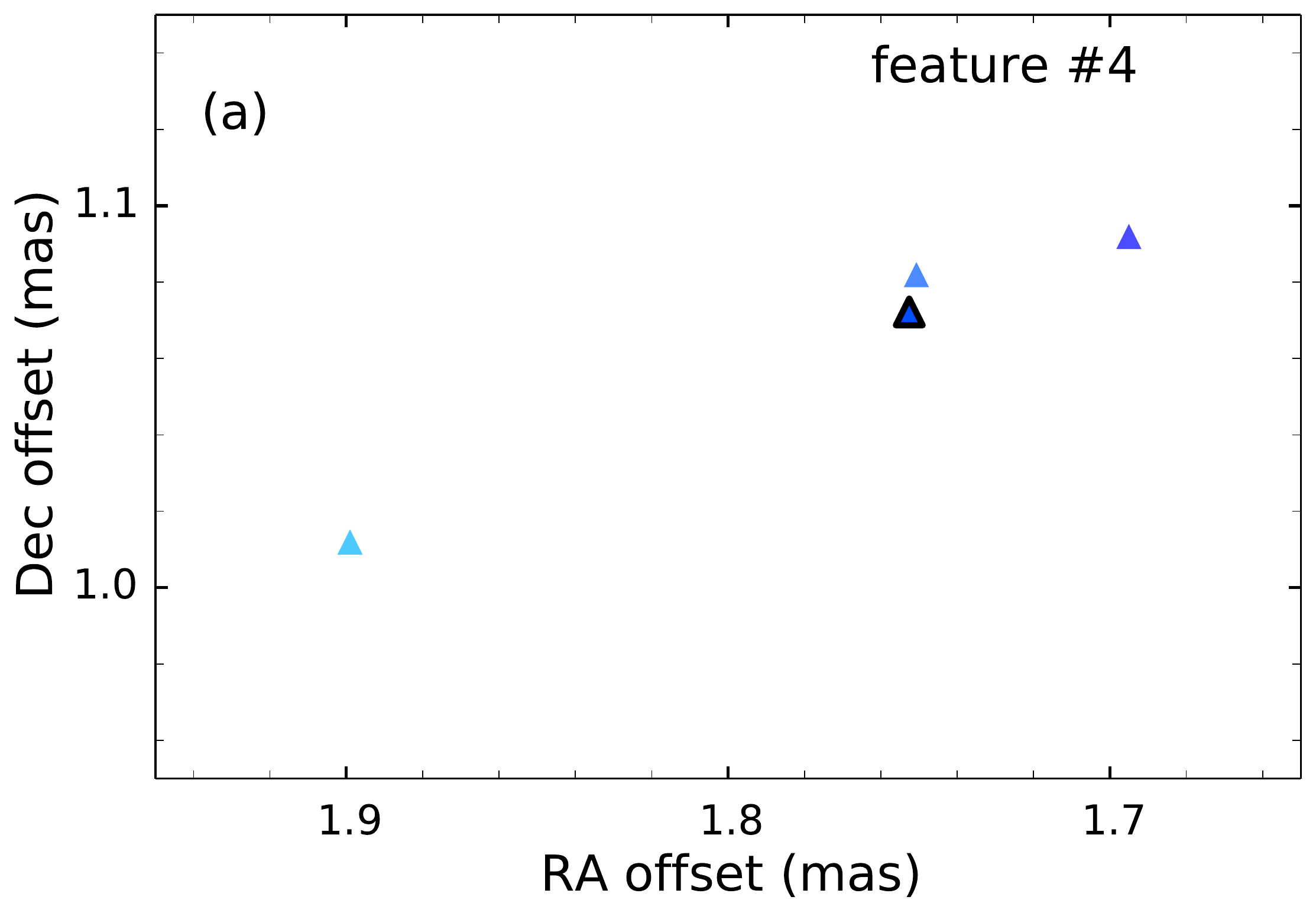}}
 {\includegraphics[width=0.43\textwidth,angle=0]{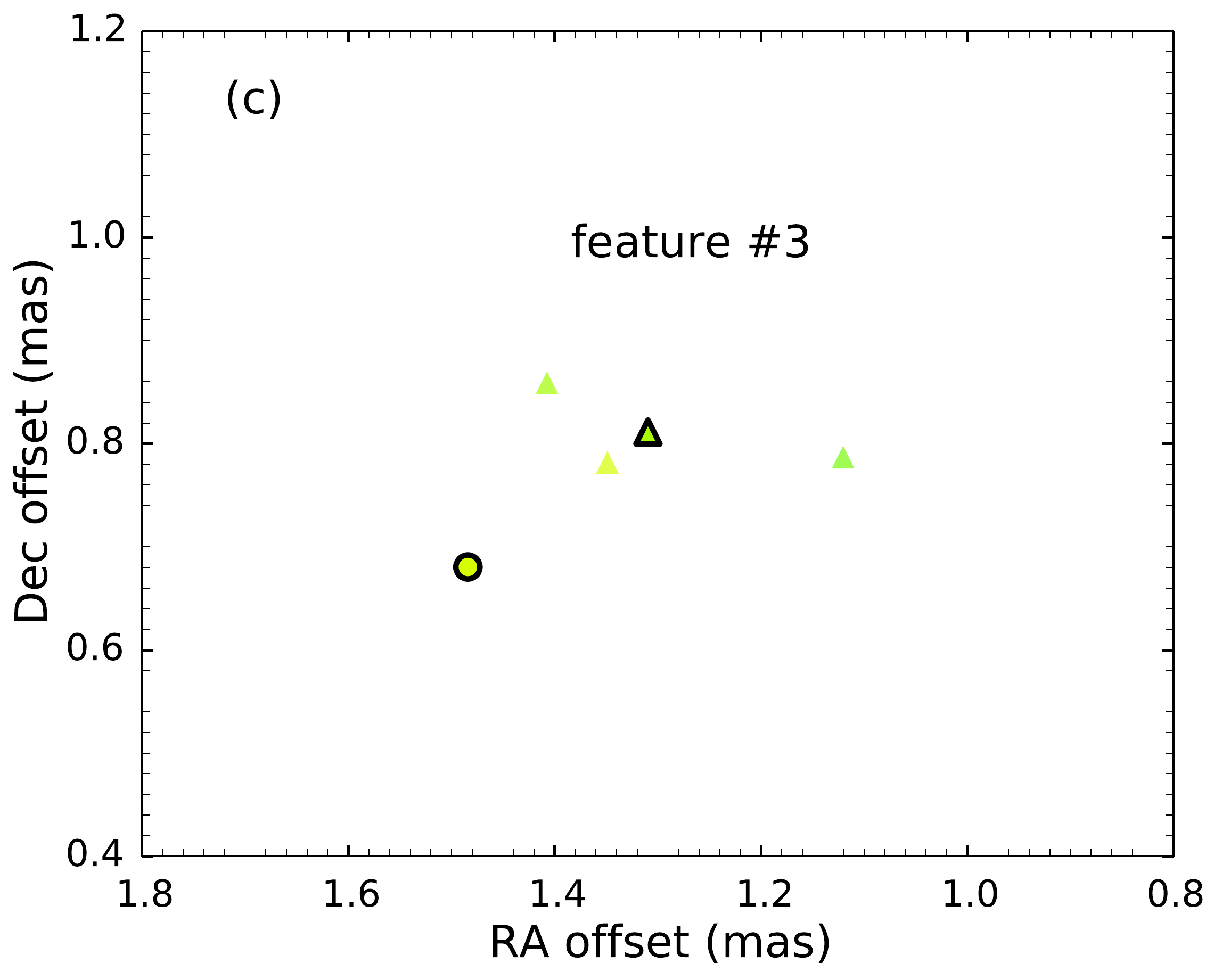}}
 {\includegraphics[width=0.38\textwidth,angle=0]{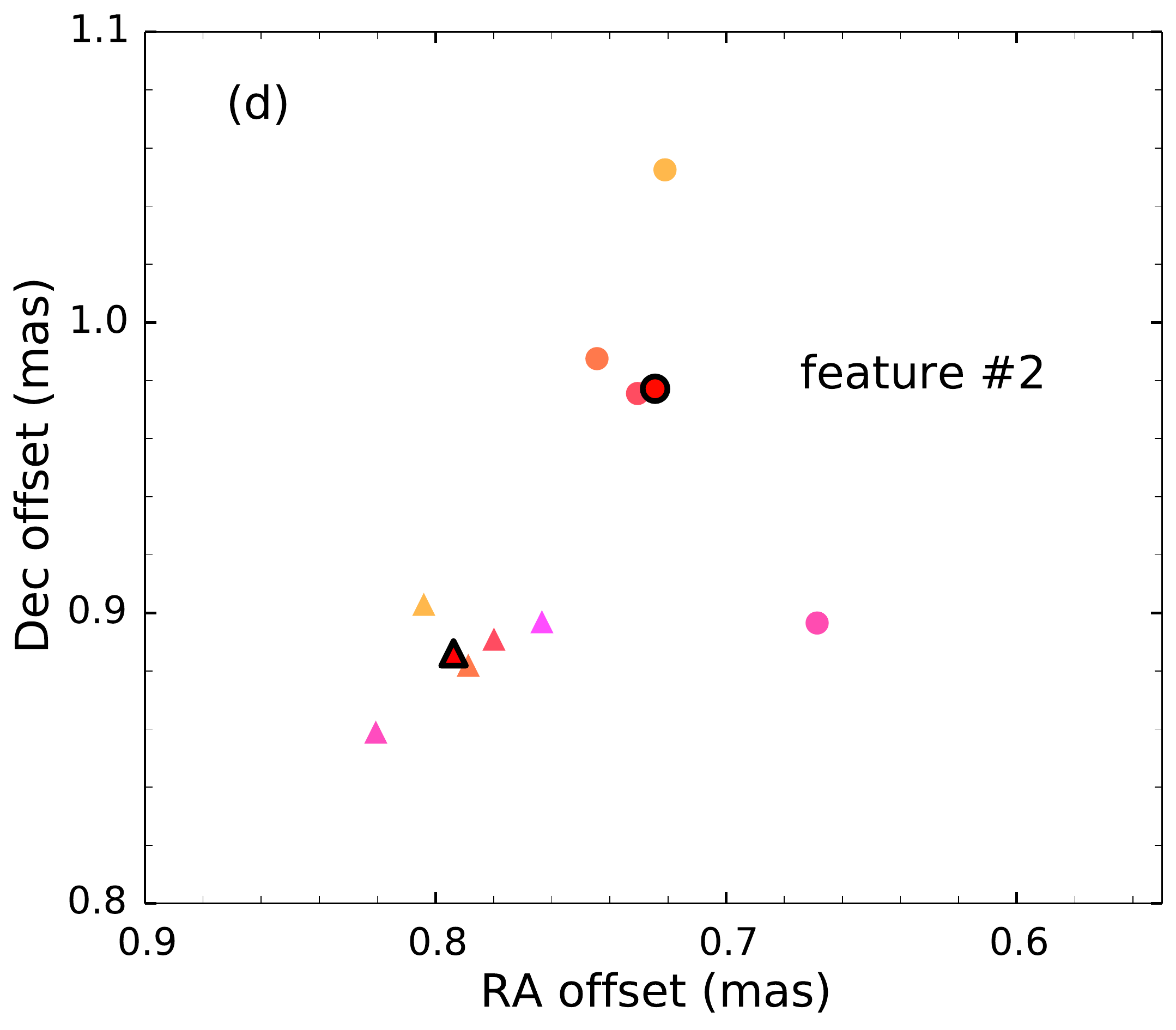}}
  {\includegraphics[width=0.46\textwidth,angle=0]{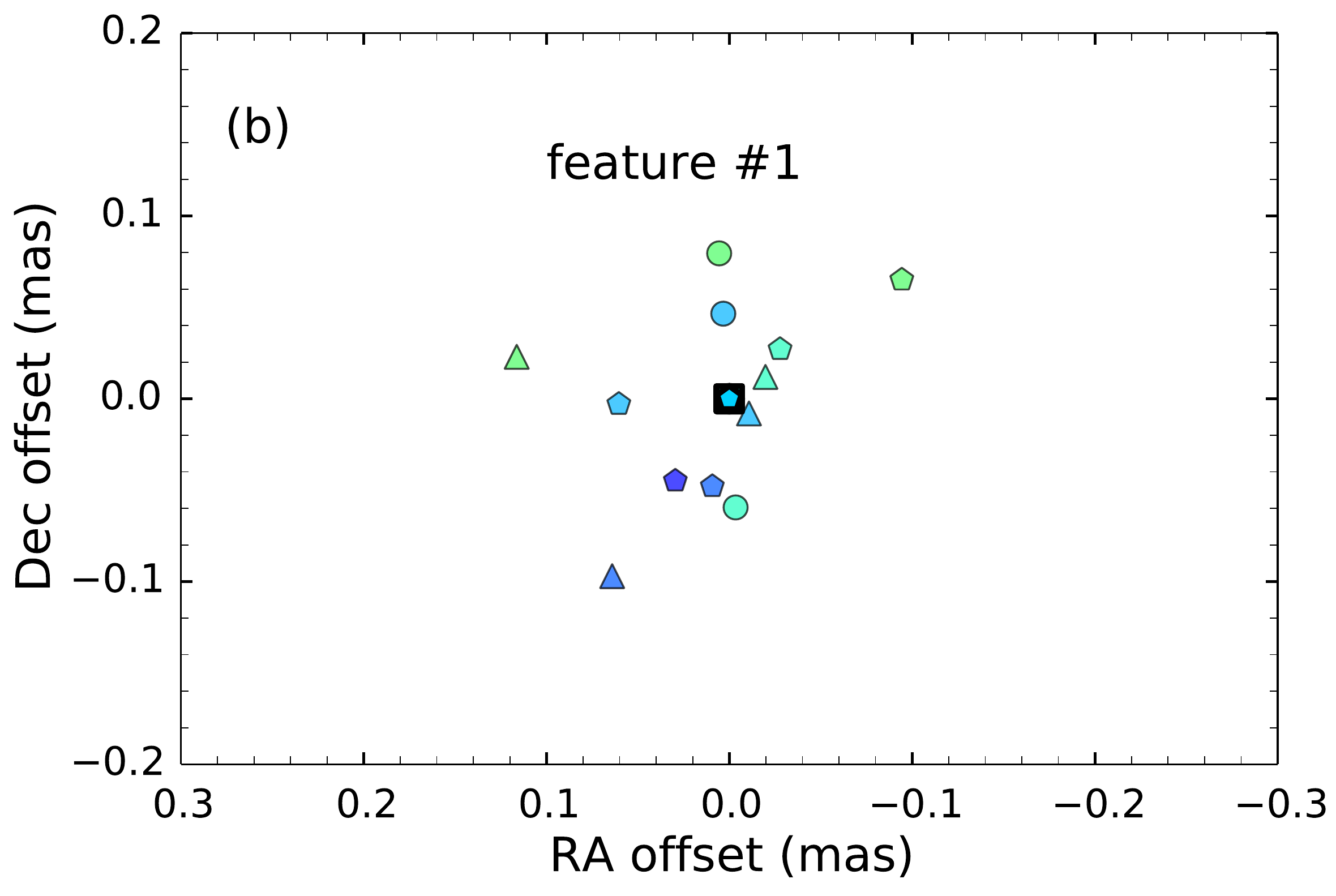}}
 {\includegraphics[width=0.6\textwidth,angle=0]{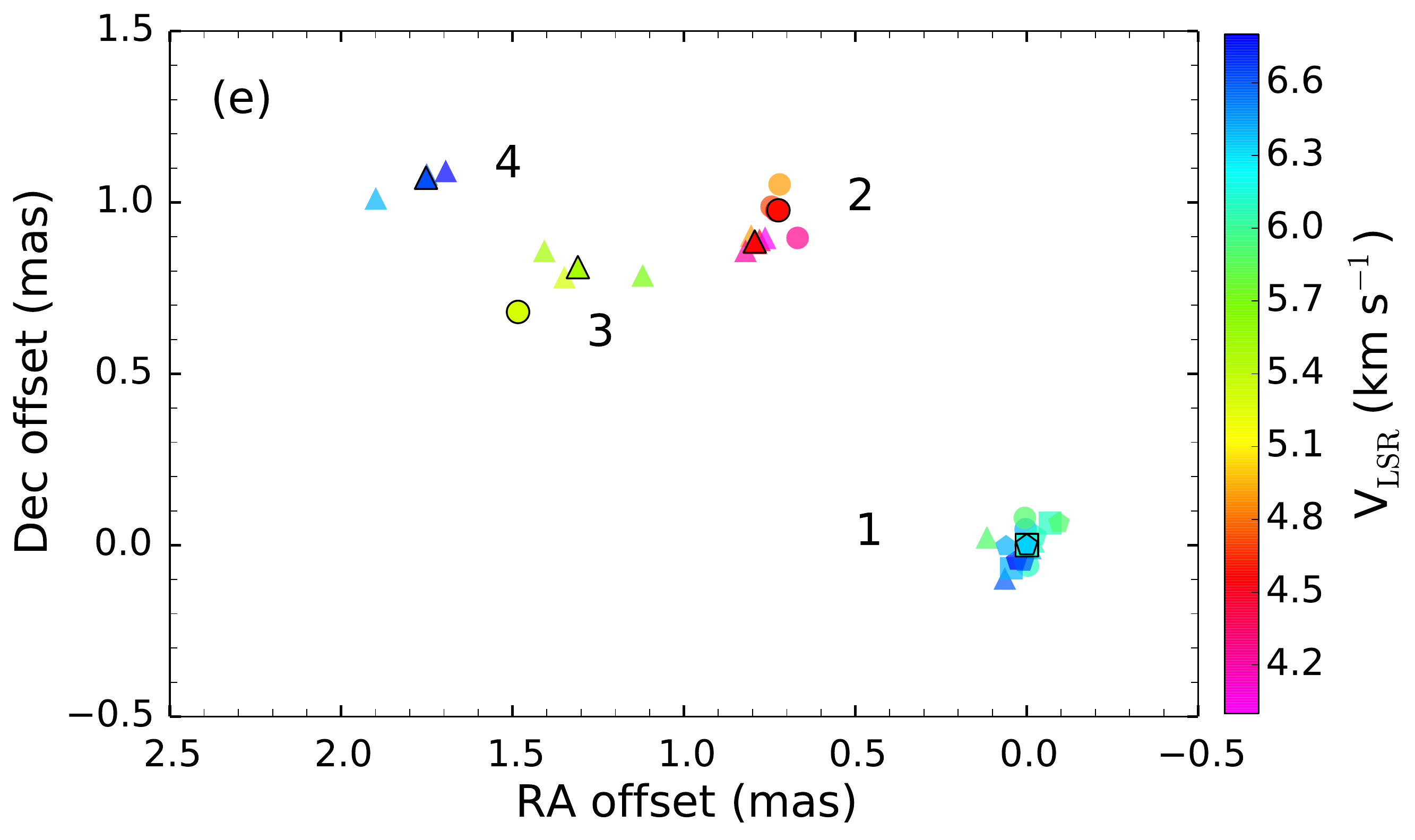}}
 \end{center}
\caption{Spatial distribution of the maser spots detected with the VLBA toward SSTgbs~J1830177--021211. The position offsets are with respect to the error-weighted mean position of feature \#1. 
%the {\it absolute} coordinates $\alpha$(J2000)=18:30:17.71415, $\delta$(J2000)=--02:12:11.6858. 
The spots are color-coded by the LSR velocity (color bar). We use different symbols to distinguish between 4 epochs observed during 2020 as follows: circles -- Mar 27, triangles -- Apr 9, squares -- Sep 29, pentagons -- Nov 1. For each epoch and feature, the symbol with black edge indicates the error-weighted mean position of all contributing spots.  Panels (a) to (d) show close-up views of the features plotted in panel (e).  
 }
\label{fig:vlba-spots}
\end{figure*}

{\bf SSTgbs J1830177--021211/IRAS~18276--0214}. The maser emission is seen at $V_{\rm LSR}$ from 4.2 to $6.6$~km~s$^{-1}$ (left panel in Fig. \ref{fig:vlba-spectra}). These velocities are blue-shifted with respect to the velocity of the cloud of $8$~km~s$^{-1}$ \citep{Kirk2013} by a few km~s$^{-1}$.  
The brightest spot has a peak flux of 0.2~Jy~beam$^{-1}$, which is higher than the highest flux detected with the VLA. Figure \ref{fig:vlba-spots} shows the spatial distribution of the VLBA detected maser spots in four epochs. We identify four main  features that occupy an extent of about 2~mas ($\approx$0.9~au) and are aligned roughly along the northeast-southwest direction. The strongest feature, labeled \#1, has persisted over the four observed epochs, which cover a time baseline of $\approx$7 months. Features \#2 and \#3 were detected on the first and second epochs, and feature \#4 only on the second epoch. 
Table \ref{tab:astro-vlba} gives the error-weighted mean position offsets and intensity-weighted $V_{\rm LSR}$ for each feature obtained from all contributing spots to that feature. These positional offsets are with respect to the position of feature \#1, which we fixed at the origin in all epochs. 
Fig. \ref{fig:vlba-spots} shows that feature \#2 (panel d) moved toward the southeast,  while feature  \#3 (panel c) moved toward the east between two consecutive epochs separated by only 13 days. Since feature positions are relative to feature \#1, we can  investigate the internal proper motions of the two features,  \#2  and  \#3. In doing this, we remove the effect of the parallax, which is not well constrained by  the current data. We obtain proper motions of $(\mu_\alpha\cos\delta,\mu_\delta)=(1.9\pm0.8,-2.6\pm1.2)$~mas~yr$^{-1}$ for feature \#2 and $(\mu_\alpha\cos\delta,\mu_\delta)=(-4.7\pm3.0,3.7\pm1.0)$~mas~yr$^{-1}$ for feature \#3. Although small, and given the fact that the positional offsets are larger than the astrometric uncertainties of about 70~$\mu$as (Sect. \ref{sec:obs-vlba}), these motions suggest that the two features are moving toward each other.
We attempt to estimate  the {\it absolute} proper motions of feature \#1 by fitting the positions of the spot detected at $V_{\rm LSR}=6.1$~km~s$^{-1}$, where the proper motions are free parameters and  the parallax is fixed to a constant value. We found that the resulting proper motions largely depend on the assumed value for the parallax. In addition, the fits yield lower residuals for parallaxes that are in the range from 0.5 to 1.0~mas.
% and the resulting motions are relative to that feature.  
%We can investigate the internal proper motions of these two features. To that end, we calculate the position offsets of the features, relative to the spot detected at $V_{\rm LSR}=6.1$~km~s$^{-1}$
%Fig. \ref{fig:vlba-spots} shows that feature \#1 (panel b) displays a non-negligible {\it absolute} motion toward the southeast, while feature  \#2 (panel d) moved toward the south between the two consecutive epochs it was detected, which were separated by only 13 days. However, we note that these motions are dominated by the contribution of the parallax, which would be of the order of 2.3~mas at the distance of Serpens South.  
Further observations spanning a larger time baseline will allow us to determine if the relative motions we measured continue over time, and disentangle {\it absolute} proper motions from the parallax.

In Fig. \ref{fig:outflows2}, we see weak blue-shifted CO emission around the location of the masers that supports the presence of a molecular outflow that is too weak to be detected. This could happen if the star is not in Serpens South, but behind the molecular cloud, which could absorb the emission from the outflow. We searched the {\it Gaia} Early Data Release 3  (EDR3) catalog and found astrometric solution for the optical counterpart of SSTgbs J1830177--021211. The parallax reported in this catalog is $1.52\pm0.84$~mas \citep{Gaia2016,GaiaEDR3}, which is still consistent (within the errors) with a distance of 436~pc, although it may suggest a larger distance. Additional observations of the maser spots will allow us to also fit the parallax and provide an independent measurement of the distance to the star. 

 It is important to note that the classification of SSTgbs J1830177--021211 as a YSO is based on the infrared spectral index \citep{Dunham_2015}. However, Asymptotic Giant Branch (AGB) stars with infrared excesses can be misidentified as YSOs and the contamination fraction is non-negligible among Class~II--Class~III sources \citep{Oliveira2009}. Thus, SSTgbs J1830177--021211 could be a background AGB star with the water masers probably tracing an expanding or contracting circumstellar envelope. Given the small relative proper motions we measured for two maser features, and the fact that smaller parallaxes are favored from the astrometric fits  and are within the $1\sigma$ uncertainty of the {\it Gaia} based parallax measurement, we incline towards the AGB star scenario as the most plausible interpretation. \\
%and the fact that smaller parallaxes are favored from astrometric fits and are within the 1\sigma uncertainty of the Gaia based parallel measurement…..

\begin{figure*}[bht]
\begin{center}
 {\includegraphics[width=0.6\textwidth,angle=0]{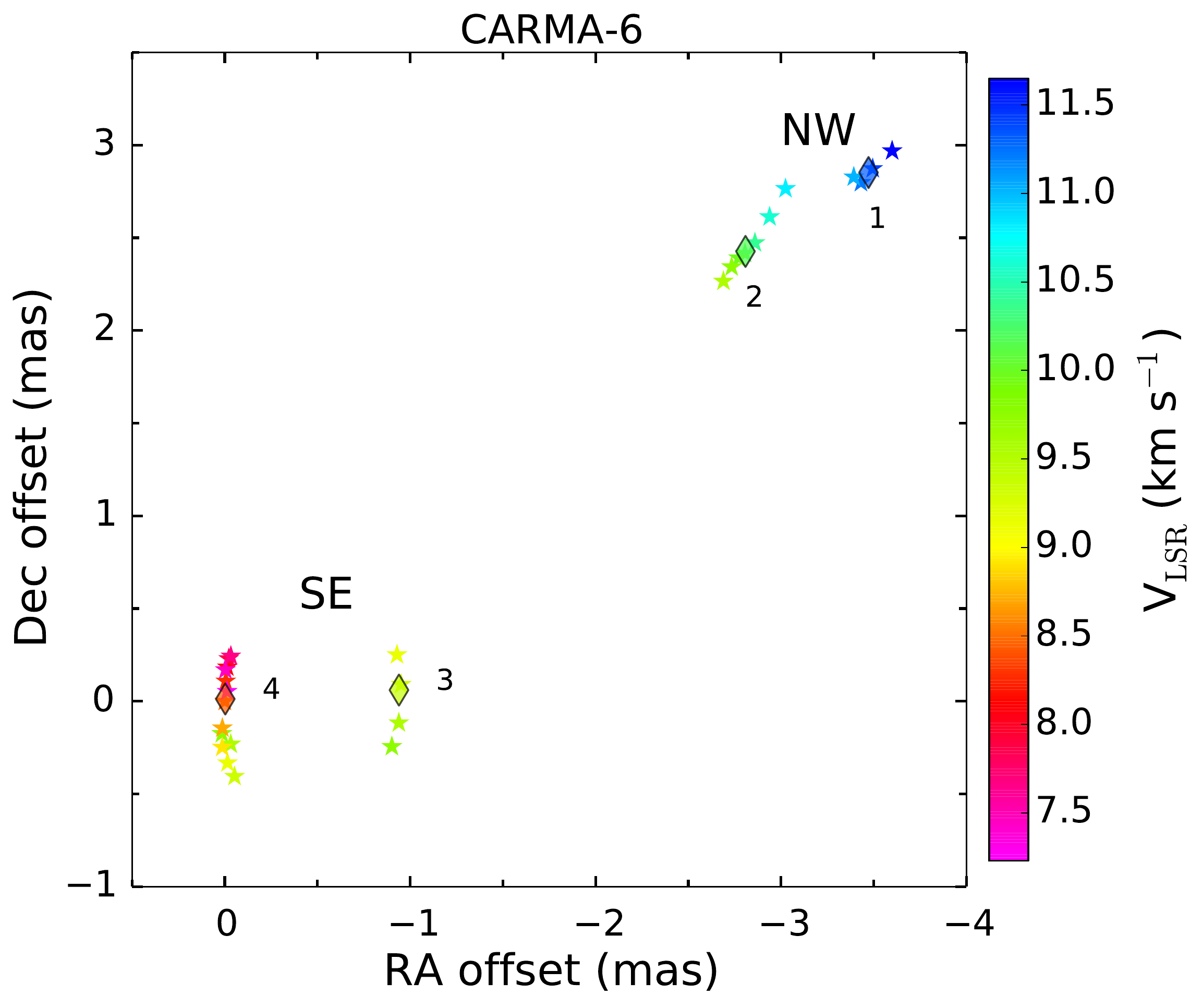}}
 \end{center}
\caption{Spatial distribution of the maser spots detected with the VLBA toward CARMA--6. The spots are color-coded by the LSR velocity (color bar). The stars indicate offsets measured on Oct 25, 2020, which are relative to $\alpha$(J2000)=18:30:03.538, $\delta$(J2000)=--02:03:08.377. For each feature, the diamonds indicate the error-weighted mean position of all contributing spots to that feature.}
\label{fig:vlba-spots-2}
\end{figure*}

{\bf CARMA-6}. Although we did not detect the maser associated with CARMA-7 using the VLBA, we did find a very bright maser  ($\sim$12~Jy~beam$^{-1}$) associated with CARMA-6. This maser was seen  serendipitously in our VLBA data on September, 2020, albeit it was not detected previously with the VLA in all three observed epochs.  Considering the rms noise level of the VLA observations (c.f. Table \ref{tab:obs}), the VLBA detection of CARMA-6 implies an increase of maser flux density by more than two orders of magnitude in the highest intensity channel. This may correspond to a flare event, although less prominent than water maser flares seen toward massive stars \citep[e.g.][]{Hirota2014,Volvach2019}. 
Additional data correlation at the position of CARMA-6 was obtained in a subsequent epoch. The spectrum observed in October, 2020 is shown in Fig \ref{fig:vlba-spectra}, after integrating over the area containing all maser spots.  
Figure \ref{fig:vlba-spots-2} shows the spatial and velocity distribution of the spots detected in the images.  
 Because the maser is very bright, in this case we phase-referenced the visibility data to the maser spot at $V_{\rm LSR}=8.5$~km~s$^{-1}$.

We detect four groups of spots or {\it features} that are oriented in the southeast-northwest direction, covering an angular extent of about 4~mas (1.7~au). %This orientation is nearly perpendicular to the orientation of the dust emission, which has a position angle of $74.8^{\rm o}$ \citep{Plunkett2018}.?
The groups located to the northwest (NW), hereafter the NW cluster, delineate a nearly straight filament. The emission is red-shifted with respect to the systemic velocity of the cloud (8~km~s$^{-1}$), covering LSR velocities smaller than the red-shifted lobe of the CO ($J=2-1$) outflow traced by ALMA at larger angular scales (Fig. \ref{fig:outflows}). We see a velocity gradient through the filament with LSR velocities increasing to the north. The groups seen to the southeast (SE), hereafter the SE cluster, show LSR velocities close to the systemic velocity. Here, the maser spots are distributed along two opposite arc-like structures, displaying velocity gradients through the arcs, with LSR velocities increasing to the south. Similar gradients have been seen for instance in Serpens SMM1 \citep[][their Figure 3]{Moscadelli2006}. In Fig. \ref{fig:vlba-spots-2}, the diamonds indicate the error-weighted mean position of all contributing emission spots (indicated by the stars) to each particular feature. The line-of-sight velocity of each feature is obtained as the intensity-weighted mean $V_{\rm LSR}$ of the contributing spots. Fig. \ref{fig:vlba-spots-2} shows that the line-of-sight velocities of the features increase to the north.  
We argue that the water masers originate in shocks between the red lobe of the molecular outflow  and the surrounding material.  As mentioned above, the NW and SE clusters draw a linear structure with the velocity gradient through this structure. The velocity gradient may arise from a rotating protostellar jet.  Observationally, rotation signatures in jets have been seen as velocity gradients  perpendicular to the jet axis \citep[e.g.][]{Chen2016,Lee2017}. In CARMA-6, the orientation of the  protostellar jet axis is not yet very well constrained. In the left panel of Fig. \ref{fig:outflows} we see that the molecular outflow is oriented close to the north-south direction, thus the jet may be oriented in the same direction. This seems to be supported by the orientation of the dust disk detected in the ALMA continuum map at 347~GHz shown in Fig. \ref{fig:alma-cont-347} of Appendix \ref{sec:appendix}. The deconvolved size of this disk is $0\rlap.{''}2\times0\rlap.{''}14$ with a position angle of 82$^{\rm o}$. If the jet is perpendicular to the disk, the jet position angle would be 172$^{\rm o}$, while the water maser filament has a  position angle of $\approx$130$^{\rm o}$. This seems to work against a rotating protostellar jet as the explanation for the observed maser velocity gradient. 

In Fig.~\ref{fig:alma-cont-347} we compare the positions of the maser spots (phase-referenced to the extragalactic calibrator) against the distribution  of the ALMA continuum emission at 347~GHz. We see that the spots are located within the disk, but have a significatn offset of 50~mas ($\approx22$~au) with respect to the continuum peak; the astrometric accuracy of the ALMA observations is about 9~mas\footnote{\url{https://help.almascience.org/kb/articles/what-is-the-astrometric-accuracy-of-alma}}. 
Because the water masers appear to locate at the base of the outflow (and within the protostellar disk), and the linear scale of the masers of 1.7~au is  smaller than the typical size of protostellar disks ($\lesssim60$~au; \citealt{Maury2019}), then the velocity gradient may inherit the velocity structure of the disk. Therefore, the observed water maser flare and the velocity gradient  may be directly linked to a disk episodic accretion burst in CARMA-6. %The spatial location of the masers suggest a jet launching region smaller than the disk size.?

The two epochs where the masers were detected are separated by only two months, covering a time baseline too short to investigate the internal kinematics of the masers. Additional VLBA observations will allow us to establish the kinematic structure of the water masers and further investigate the above alternative scenarios.

\subsection{Continuum sources detected with the VLA}\label{sec:cont}

We performed a visual inspection of the maps that were constructed for the 48 VLA fields, first looking at the individual epochs, and then at the maps of the combination of the data from three epochs (see Sec. \ref{sec:vlaobs}). The visual inspection was done in the images uncorrected for the primary beam response, as this correction increases the noise toward the field edges affecting weak sources that then may appear as noise. However, once identified, the properties of the sources are measured in the primary beam corrected images.
Maps of $9''\times9''$ in size around the location of detected sources are presented in Figures \ref{fig:vla-continuum-all}--\ref{fig:vla-continuum-bck} in Appendix \ref{sec:appendix}. The maps are for all available epochs, but we note that some epochs do not exhibit detection. Table \ref{tab:imfit} lists the 17 sources detected with radio continuum, 
as well as their positions and fluxes as obtained by fitting the brightness distribution with a Gaussian model using the task {\tt imfit} in CASA. 
The fluxes are listed for each epoch and for the combined image.

Not all detected sources with radio continuum are associated with known young stars or other type of objects; there are 5 sources that have no counterparts (within a radius of 2$''$) in SIMBAD\footnote{\url{http://simbad.u-strasbg.fr/simbad/}}. On the other hand, we found that 10 sources are associated with known or candidate YSOs \citep{Povich2013}, and other 2 are associated with known radio sources \citep{Condon1998,OrtizLeon2015,Kern2016AJ}, also within a radius of 2$''$. Table \ref{tab:imfit} gives the names of the known sources. Out of the 12 objects that have an association with a known source, 6 had not been detected before in the radio according to SIMBAD. Therefore, we are reporting 6+5=11 new radio continuum detections. 

The newly detected radio continuum sources with no counterparts at any other wavelength are \#1, \#5, \#14, \#15 and \#17. Source \#1 is detected in the four observed epochs with fluxes of 1.7--1.9~mJy. The other sources (\#5, \#14, \#15 and \#17) are detected in only one epoch, with fluxes above 0.22~mJy. In addition, sources  \#3 and  \#12, that have reported before in the literature, do not have counterparts at any other wavelength as well.

Following \cite{Anglada1998}, we can estimate the number of expected background sources inside a field of diameter $\theta_F$ as,

\begin{equation}
%\begin{split}
    N  = 1.4 \left\lbrace 1 -\exp \left[  -0.0066 \left( \frac{\theta_F}{\rm arcmin} \right)^2  \left( \frac{\nu}{\rm 5~GHz} \right)^2  \right] \right\rbrace   \times  \left( \frac{S_0}{\rm mJy} \right)^{-0.75} \left( \frac{\nu}{\rm 5~GHz} \right)^{-2.52}
%\end{split}
\end{equation}

\noindent where $S_0$ is the detectable flux density threshold and $\nu$ the observing frequency. In our observations, $\nu$=22.2~GHz, and $S_0=3\times{\rm rms}\approx0.09$~mJy (c.f.\ Sect. \ref{sec:vlaobs}). Using a field size of $\theta_F=2\rlap.{'}7$, we obtain $\approx7$ expected background objects in the 48 observed fields. Thus, all of the unclassified sources with detected radio continuum emission are probably extragalactic objects. 

Since our targets were observed in multiple epochs, covering a timescale of about 3 weeks, we can investigate the variability of continuum emission between the epochs. We estimated the variability as the difference between the maximum and minimum peak flux density, normalized by the maximum flux. For the estimation of variability uncertainties, we adopted a flux density calibration error of 15\%\footnote{\url{https://science.nrao.edu/facilities/vla/docs/manuals/oss/performance/fdscale}}, which was added quadratically to the statistical errors obtained from the Gaussian fits. We found that 9 sources show high levels of variability, with variations $\gtrsim50\%$ at $3\sigma$. These sources are \#4, \#5, \#11, \#12, \#13, \#14, \#15, \#16 and \#17. Four of these objects are YSOs; the other 5 are background candidates. Thus, in terms of variability, we do not see a distinction between the two groups. Previous works have found a similar result at shorter radio wavelengths. For instance, \cite{Kounkel2017} showed that both YSOs and extragalactic objects show strong radio continuum variability at 7.5~GHz.

\section{Discussion}\label{sec:discussion}

The four sources with H$_2$O maser emission detected here are known to be associated with phenomena related to YSOs. However, while CARMA-7, CARMA-6 and SSTgbs J1829053--014156 are in the early Class 0--Class I phase, SSTgbs J1830177--021211 is probably  in the more evolved Class II phase. Three of the sources with associated maser emission drive large-scale outflows. From the spatial distribution of the maser spots, we argue that in all these sources the masers originate very close to the star, and are excited by the interaction between molecular outflows with the surrounding dense material, likely of the circumstellar disk.

\begin{figure}[!bht]
\begin{center}
 {\includegraphics[width=0.4\textwidth,angle=0]{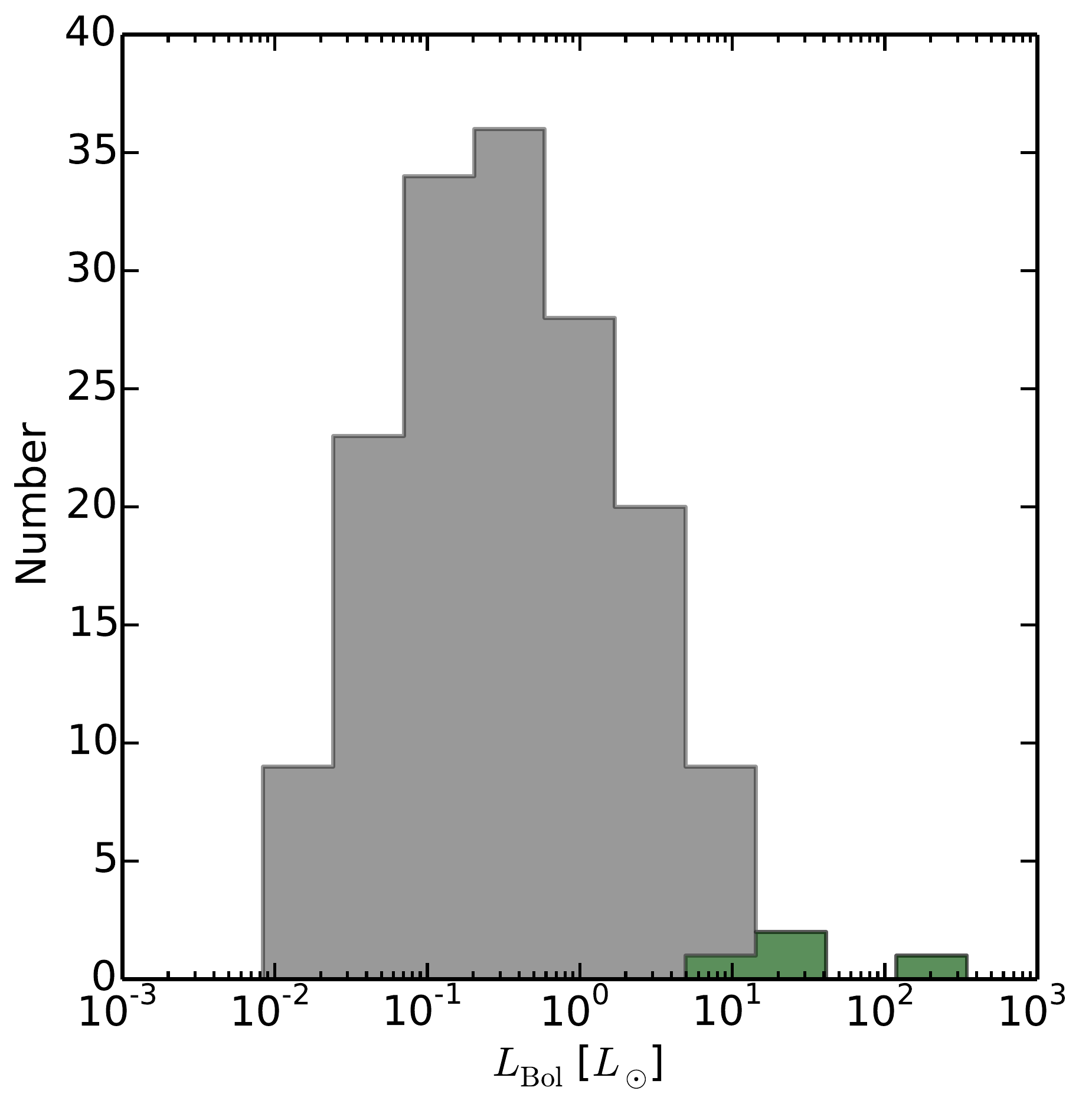}}
  {\includegraphics[width=0.5\textwidth,angle=0]{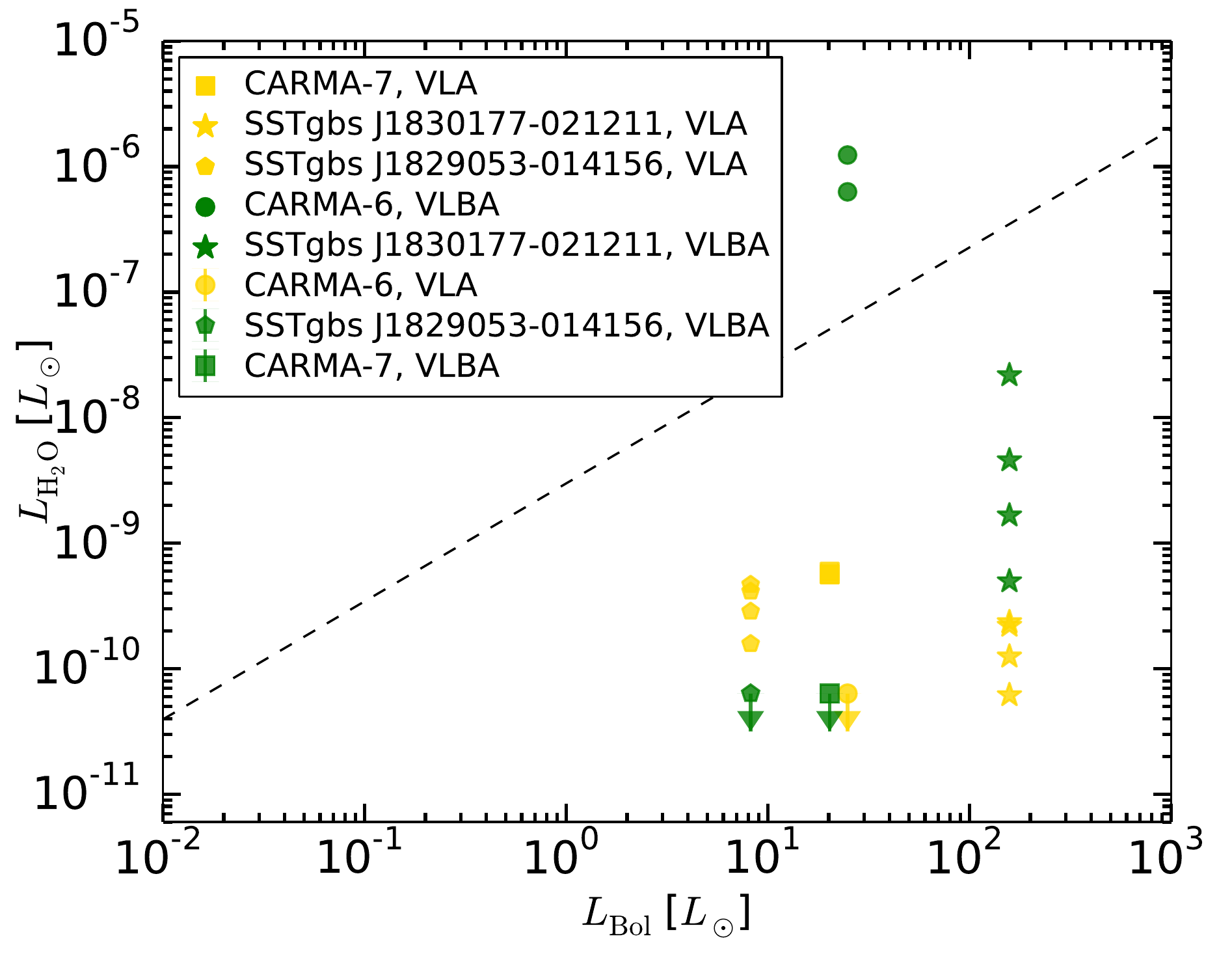}}
 \end{center}
\caption{{\it Left}: Bolometric luminosity distribution of 162 YSOs covered by our VLA observations in grey and the 4 objects with detected water maser emission. {\it Right}: Water maser luminosity versus bolometric luminosity for the objects detected with the VLA and the VLBA as described in the legend. The arrows indicate upper limits. The dashed line represents the empirical relation expressed by Eq. \ref{eq:Shirley}  \citep{Shirley2007}.
}
\label{fig:lbol}
\end{figure}

 Extinction corrected bolometric luminosities are available for the 162 stars of the catalog by \cite{Dunham_2015} that were observed with the VLA. The distribution of the bolometric luminosities, which have been rescaled assuming a distance of 436~pc, are shown in the left panel of Fig. \ref{fig:lbol}.  Also shown in this figure are the bolometric luminosities of SSTgbs J1829053--014156, and SSTgbs J1830177--021211, and the internal luminosities of CARMA-7 and CARMA-6. As expected, maser emission was detected toward the objects with the highest luminosity. Figure \ref{fig:lbol} suggests that there is a bolometric luminosity threshold of $L_{\rm Bol}\approx10~L_\odot$ to excite water maser emission. However, water masers have been detected in objects with lower luminosities before \citep{Furuya2003}, for example, in VLA~1623  ($L_{\rm Bol}\approx1~L_\odot$, $d$ = 138~pc; \citealt{Andre1993,Ortiz2018}) and GF 9-2 ($L_{\rm Bol}\approx0.3-1.7~L_\odot$, $d$ = 200 -- 474~pc; \citealt{Furuya2008,Podio2020}). 
 It is still possible that the detection of water masers associated to lower-luminosity objects in Serpens South was missed due to variability. For instance, in CARMA-6, the masers were not detected in the three observed epochs with the VLA, but serendipitously detected with the VLBA about 1.5 years later. 

We estimate water maser luminosities according to
\begin{equation}\label{eq:lmaser}
L_{\rm H_2O} = 4\pi d^2 S_{\rm int} \Delta V \nu_0/c, 
\end{equation}
\noindent where $S_{\rm int}$ is the maser integrated flux density, $\Delta V$ is the velocity range of maser emission, $\nu_0=22,235.080$~MHz is the rest frequency of the $J=6_{1,6}-5_{2,3}$ water line, $c$ the speed of light, and $d$ the distance to the source. {The water mater luminosities are listed in Column 9 of Tables \ref{tab:line-imfit} and \ref{tab:astro-vlba} for sources detected with the VLA and VLBA, respectively. }
Using a 3$\sigma$ channel sensitivity of $\approx$48~mJy from our VLA observations (c.f.\ Table \ref{tab:obs}), a velocity spread of the masers of 3 channels, and $d=436$~pc, Eq. \ref{eq:lmaser} gives
an H$_2$O luminosity of $6\times10^{-11}~L_\odot$. Assuming the correlation between $L_{\rm H_2O}$ and $L_{\rm Bol}$ found by \cite{Shirley2007} for high-luminosity YSOs, according to which 
\begin{equation}\label{eq:Shirley}
L_{\rm H_2O} = 3\times10^{-9}L_{\rm Bol}^{0.94},
\end{equation}
 \noindent this upper limit in H$_2$O luminosity corresponds to $L_{\rm Bol}\approx0.02~L_\odot$.  Thus, our VLA observations were in principle sensitive enough to detect all H$_2$O masers associated to low-luminosity protostars in Serpens South with $L_{\rm Bol}\gtrsim0.02~L_\odot$.  As noted by \cite{Gomez2017}, the correlation between bolometric and maser luminosities may not hold for the lowest-luminosity YSOs. We plot in the right panel of Fig. \ref{fig:lbol} the bolometric luminosities of the 4 objects that have water masers and the maser luminosities measured at each individual epoch observed with the VLA and the VLBA. Due to the strong variability in both flux and velocity spread of the maser emission, the $L_{\rm H_2O}$ changes in all sources by more than 1 order of magnitude. CARMA-6 shows the highest variability, since the non-detection with the VLA implies a change in $L_{\rm H_2O}$ by about 4 orders of magnitude. In Fig. \ref{fig:lbol}, the two stars detected with the VLBA (SSTgbs~J1830177--021211 and CARMA-6) fall, within one order of magnitude, close to their predicted position by the $L_{\rm H_2O}$ versus $L_{\rm Bol}$ empirical relationship. A scatter of one order of magnitude was also observed for this relationship \cite[][their Fig.\ 3]{Shirley2007}.

\section{Conclusions}\label{sec:conclusions}

We have conducted an interferometric survey of 22~GHz H$_2$O masers toward the low-mass star-forming region Serpens South. Our observations were first carried out with the VLA covering all known protostars (Class 0--Class I objects)  across the region.  The VLA observations revealed, for the first time, three water masers in the region, which are found to be associated to CARMA-7, SSTgbs J1830177--021211 and SSTgbs J1829053--014156. Follow-up VLBA  observations were carried out toward the VLA-detected sources to investigate the spatial distribution and kinematics of the masers. The VLBA observations found water maser emission associated to  CARMA-6, which had not been detected with the VLA. 

Three water maser sources (CARMA-7, SSTgbs J1829053--014156 and CARMA-6) are associated with Class 0--Class I objects that drive large scale molecular outflows and also display radio continuum emission from ionized gas. The water masers are found at the base of the molecular outflows and we propose that in all these three objects the masers are excited in shocks driven by the interaction between a protostellar jet and the circumstellar material. On the other hand, the  source responsible for the excitation of the water maser associated with SSTgbs J1830177--021211 is unknown.  This source has been classified in the literature as a Class II object and has no associated molecular outflows or radio jets. The small relative proper motions of two maser features that persisted over two epochs, and the small parallax hinted by the astrometric fits to the brightest feature suggest that SSTgbs J1830177--021211 is most likely a background AGB star with the water masers tracing an expanding or contracting circumstellar envelope. Further VLBI observations will allow us to obtain the parallax and proper motions of the maser spots and to test the proposed mechanism for the water maser excitation in these objects and confirm the  AGB scenario proposed for SSTgbs J1830177--021211.

We also investigate the distributions of the bolometric luminosity of sources hosting 22~GHz H$_2$O masers and 162 YSOs covered by our observations. The comparison of the two distributions suggest a luminosity threshold for the water maser emission of $L_{\rm Bol}\approx10~L_\odot$. However, the water masers show strong variability, thus lower-luminosity sources may have been missed by the observations.

Lastly, we detected 11 new sources with radio continuum emission at~22 GHz, of which 6 are known or candidate YSOs, and 5 are unknown sources without counterparts at any other wavelength. Based on the estimation of the number of expected background sources in the observed area,  we suggest that all of these unclassified sources are probably extragalactic objects.

%__________________________________________________________________
%__________________________________________________________________

\acknowledgments

The authors are grateful to the anonymous referee, whose comments helped to improve this paper. 
G.N.O.-L.\ acknowledges support from the von Humboldt Stiftung.
L.L.\ acknowledges the support of DGAPA/PAPIIT grants IN112417 and IN112820, CONACyT-AEM grant 275201, and CONACyT-CF grant 263356. The authors acknowledge MiaoMiao Zhang for sharing his Canada--France--Hawaii
Telescope near-infrared data. The National Radio Astronomy Observatory is a facility of the National Science Foundation operated under a cooperative agreement by Associated Universities, Inc. This paper makes use of the following ALMA data: ADS/JAO.ALMA \#2012.1.00769.S and \#2015.1.00283.S. ALMA is a partnership of ESO (representing its member states), NSF (USA) and NINS (Japan), together with NRC (Canada), MOST and ASIAA (Taiwan), and KASI (Republic of Korea), in cooperation with the Republic of Chile. The Joint ALMA Observatory is operated by ESO, AUI/NRAO and NAOJ.

\vspace{5mm}
%\facilities{HST(STIS), Swift(XRT and UVOT), AAVSO, CTIO:1.3m,
%CTIO:1.5m,CXO}

%% Similar to \facility{}, there is the optional \software command to allow 
%% authors a place to specify which programs were used during the creation of 
%% the manuscript. Authors should list each code and include either a
%% citation or url to the code inside ()s when available.

%\software{astropy \citep{2013A&A...558A..33A}
%          Cloudy \citep{2013RMxAA..49..137F}, 
%          SExtractor \citep{1996A&AS..117..393B}
 %        }

%% Appendix material should be preceded with a single \appendix command.
%% There should be a \section command for each appendix. Mark appendix
%% subsections with the same markup you use in the main body of the paper.

%% Each Appendix (indicated with \section) will be lettered A, B, C, etc.
%% The equation counter will reset when it encounters the \appendix
%% command and will number appendix equations (A1), (A2), etc. The
%% Figure and Table counter will not reset.

%%\begin{center}
%\begin{landscape}
%\begin{longtable}{ccccccccc }
%\caption{Properties of the VLA detected radio continuum sources.}
%\label{tab:imfit} \\
%\hline
%\hline  
%Source  & Epoch & $\alpha$  & $\delta$   & Peak Flux   & Int.\ Flux    &   Known & Object\tablefootmark{a} & Ref.\tablefootmark{a}   \\
%ID          &            & (J2000)  & (J2000)   &(mJy~beam$^{-1}$)  &  (mJy)   &    Name & Type     \\
%\hline 
%\endfirsthead
%\caption{\it Continued.}\\
%\hline\hline
%Source  & Epoch & $\alpha$  & $\delta$   & Peak Flux   & Int.\ Flux    &  Known & Object\tablefootmark{a}  & Ref.\tablefootmark{a}  \\
%ID          &            & (J2000)  & (J2000)   &(mJy~beam$^{-1}$)  &  (mJy)   &   Name & Type     \\
%\hline
%\endhead
%\hline
%\endfoot
%\hline
%\endlastfoot
%
\begin{longrotatetable}
\begin{deluxetable*}{cccccccccc}
\tablecaption{Properties of the VLA detected radio continuum sources. \label{tab:imfit}}
\tablewidth{700pt}
\tabletypesize{\scriptsize}
\tablehead{
\colhead{Source} & \colhead{Epoch} & 
\colhead{$\alpha$ (J2000)} & \colhead{$\delta$ (J2000)} & 
\colhead{Peak Flux} & \colhead{Int.\ Flux} & 
\colhead{Known} & \colhead{Object$^{a}$} &  \colhead{Ref.$^{a}$} & \colhead{New$^{b}$} \\ 
\colhead{ID} & \colhead{} & \colhead{(h:m:s)} & \colhead{($^{\rm o}$:$'$:$''$)} & 
\colhead{(mJy~beam$^{-1}$)} & \colhead{ (mJy)} & \colhead{Name} &
\colhead{Type} &  & \colhead{Detection?} \\
   \colhead{(1)} & \colhead{(2)} & \colhead{(3)} & \colhead{(4)} & \colhead{(5)} & \colhead{(6)} & 
 \colhead{(7)} & \colhead{(8)} & \colhead{(9)} & \colhead{(10)}
} 
\startdata
1  &  1  &  18:29:04.88  &  -01:30:06.2  &  1.75 $\pm$   0.02  &  1.78 $\pm$   0.03  &    &    &    &  \multirow{5}{*}{Y}  \\ 
1  &  2  &  18:29:04.88  &  -01:30:06.1  &  1.63 $\pm$   0.02  &  1.67 $\pm$   0.04  &    &    &    &    \\ 
1  &  3  &  18:29:04.89  &  -01:30:06.1  &  1.91 $\pm$   0.04  &  1.84 $\pm$   0.07  &    &    &    &    \\ 
1  &  4  &  18:29:04.88  &  -01:30:06.2  &  1.89 $\pm$   0.05  &  1.93 $\pm$   0.10  &    &    &    &    \\ 
1  &  all  &  18:29:04.88  &  -01:30:06.2  &  1.85 $\pm$   0.02  &  1.89 $\pm$   0.04  &    &    &    &    \\ 
\hline
2  &  1  &  18:29:05.33  &  -01:41:57.0  &  0.75 $\pm$   0.02  &  0.79 $\pm$   0.03  &  \multirow{5}{*}{IRAS 18264-0143}  &  \multirow{5}{*}{Y*O}  &  \multirow{5}{*}{1}  &  \multirow{5}{*}{Y, H$_2$O}  \\ 
2  &  2  &  18:29:05.33  &  -01:41:57.0  &  0.73 $\pm$   0.02  &  0.78 $\pm$   0.03  &    &    &    &    \\ 
2  &  3  &  18:29:05.32  &  -01:41:56.9  &  0.72 $\pm$   0.04  &  0.79 $\pm$   0.07  &    &    &    &    \\ 
2  &  4  &  18:29:05.33  &  -01:41:57.0  &  0.71 $\pm$   0.02  &  0.78 $\pm$   0.04  &    &    &    &    \\ 
2  &  all  &  18:29:05.33  &  -01:41:57.0  &  0.80 $\pm$   0.01  &  0.91 $\pm$   0.03  &    &    &    &    \\ 
\hline
3  &  1  &  18:29:06.29  &  -02:07:48.1  &  6.94 $\pm$   0.29  &  6.78 $\pm$   0.51  &  \multirow{5}{*}{PMN J1829-0207}  &  \multirow{5}{*}{Rad}  &  \multirow{5}{*}{2}  &    \\ 
3  &  2  &  18:29:06.29  &  -02:07:48.1  &  8.45 $\pm$   0.34  &  9.89 $\pm$   0.67  &    &    &    &    \\ 
3  &  3  &  18:29:06.29  &  -02:07:48.2  &  8.51 $\pm$   0.28  &  9.05 $\pm$   0.52  &    &    &    &    \\ 
3  &  4  &  18:29:06.29  &  -02:07:48.1  &  7.07 $\pm$   0.09  &  7.34 $\pm$   0.17  &    &    &    &    \\ 
3  &  all  &  18:29:06.29  &  -02:07:48.1  &  8.03 $\pm$   0.15  &  8.34 $\pm$   0.28  &    &    &    &    \\ 
\hline
4  &  1  &  --  &  --  &  $<$0.34  &  --  &  \multirow{4}{*}{2MASS J18291560-0204503}  &  \multirow{4}{*}{Y*O}  &  \multirow{4}{*}{3}  &  \multirow{4}{*}{Y}  \\ 
4  &  2  &  18:29:15.62  &  -02:04:50.3  &  2.75 $\pm$   0.06  &  2.82 $\pm$   0.10  &    &    &    &    \\ 
4  &  4  &  --  &  --  &  $<$0.18  &  --  &    &    &    &    \\ 
4  &  all  &  18:29:15.62  &  -02:04:50.3  &  1.15 $\pm$   0.04  &  1.42 $\pm$   0.08  &    &    &    &    \\ 
\hline
5  &  1  &  18:29:51.04  &  -01:54:24.4  &  0.22 $\pm$   0.01  &  0.18 $\pm$   0.02  &    &    &    &  \multirow{4}{*}{Y}  \\ 
5  &  2  &  --  &  --  &  $<$0.06  &  --  &    &    &    &    \\ 
5  &  4  &  --  &  --  &  $<$0.06  &  --  &    &    &    &    \\ 
5  &  all  &  --  &  --  &  $<$0.06  &  --  &    &    &    &    \\ 
\hline
6  &  1  &  18:30:01.36  &  -02:10:25.7  &  0.32 $\pm$   0.01  &  0.33 $\pm$   0.03  &  \multirow{4}{*}{2MASS J18300136-0210256, G028.5480+03.7663}  &  \multirow{4}{*}{Y*O}  &  \multirow{4}{*}{1}  &  \multirow{4}{*}{Y}  \\ 
6  &  2  &  18:30:01.36  &  -02:10:25.7  &  0.37 $\pm$   0.01  &  0.36 $\pm$   0.03  &    &    &    &    \\ 
6  &  4  &  18:30:01.36  &  -02:10:25.7  &  0.32 $\pm$   0.01  &  0.31 $\pm$   0.02  &    &    &    &    \\ 
6  &  all  &  18:30:01.35  &  -02:10:25.7  &  0.38 $\pm$   0.01  &  0.38 $\pm$   0.02  &    &    &    &    \\ 
\hline
7  &  1  &  18:30:03.12  &  -01:36:32.9  &  0.16 $\pm$   0.02  &  0.16 $\pm$   0.03  &  \multirow{5}{*}{G029.0540+04.0193}  &  \multirow{5}{*}{Y*?}  &  \multirow{5}{*}{4}  &  \multirow{5}{*}{Y}  \\ 
7  &  2  &  18:30:03.12  &  -01:36:33.1  &  0.15 $\pm$   0.01  &  0.18 $\pm$   0.03  &    &    &    &    \\ 
7  &  3  &  --  &  --  &  $<$0.10  &  --  &    &    &    &    \\ 
7  &  4  &  --  &  --  &  $<$0.09  &  --  &    &    &    &    \\ 
7  &  all  &  18:30:03.12  &  -01:36:32.9  &  0.19 $\pm$   0.01  &  0.20 $\pm$   0.01  &    &    &    &    \\ 
\hline
8  &  1  &  --  &  --  &  $<$0.10  &  --  &  \multirow{4}{*}{MHO 3247, G028.6658+03.8174, [KKT2016] VLA 11}  &  \multirow{4}{*}{Y*O, Rad}  &  \multirow{4}{*}{5}  &    \\ 
8  &  2  &  18:30:03.38  &  -02:02:45.8  &  0.15 $\pm$   0.01  &  0.13 $\pm$   0.02  &    &    &    &    \\ 
8  &  4  &  --  &  --  &  $<$0.08  &  --  &    &    &    &    \\ 
8  &  all  &  18:30:03.37  &  -02:02:45.8  &  0.19 $\pm$   0.02  &  0.23 $\pm$   0.03  &    &    &    &    \\ 
\hline
9  &  1  &  18:30:03.54  &  -02:03:08.4  &  0.63 $\pm$   0.03  &  0.55 $\pm$   0.05  &  \multirow{4}{*}{SSTYSV J183003.48-020308.5, CARMA-6, [KKT2016] VLA 13}  &  \multirow{4}{*}{Y*O, Rad}  &  \multirow{4}{*}{3}  &  \multirow{4}{*}{ H$_2$O}  \\ 
9  &  2  &  18:30:03.54  &  -02:03:08.3  &  0.73 $\pm$   0.02  &  0.71 $\pm$   0.04  &    &    &    &    \\ 
9  &  4  &  18:30:03.54  &  -02:03:08.4  &  0.56 $\pm$   0.02  &  0.54 $\pm$   0.03  &    &    &    &    \\ 
9  &  all  &  18:30:03.54  &  -02:03:08.4  &  0.77 $\pm$   0.02  &  0.72 $\pm$   0.04  &    &    &    &    \\ 
\hline
10  &  1  &  18:30:04.11  &  -02:03:02.5  &  0.29 $\pm$   0.01  &  0.44 $\pm$   0.03  &  \multirow{4}{*}{CARMA-7, [KKT2016] VLA 12}  &  \multirow{4}{*}{Y*O, Rad}  &  \multirow{4}{*}{1}  &  \multirow{4}{*}{ H$_2$O}  \\ 
10  &  2  &  18:30:04.12  &  -02:03:02.7  &  0.35 $\pm$   0.02  &  0.40 $\pm$   0.03  &    &    &    &    \\ 
10  &  4  &  18:30:04.13  &  -02:03:02.6  &  0.28 $\pm$   0.01  &  0.37 $\pm$   0.03  &    &    &    &    \\ 
10  &  all  &  18:30:04.12  &  -02:03:02.6  &  0.38 $\pm$   0.02  &  0.51 $\pm$   0.04  &    &    &    &    \\ 
\hline
11  &  1  &  --  &  --  &  $<$0.09  &  --  &  \multirow{5}{*}{2MASS J18300580-0201444, [KKT2016] VLA 7}  &  \multirow{5}{*}{Y*O, Rad}  &  \multirow{5}{*}{5}  &    \\ 
11  &  2  &  --  &  --  &  $<$0.11  &  --  &    &    &    &    \\ 
11  &  3  &  18:30:05.82  &  -02:01:44.6  &  0.23 $\pm$   0.01  &  0.21 $\pm$   0.02  &    &    &    &    \\ 
11  &  4  &  --  &  --  &  $<$0.11  &  --  &    &    &    &    \\ 
11  &  all  &  18:30:05.82  &  -02:01:44.4  &  0.21 $\pm$   0.01  &  0.28 $\pm$   0.03  &    &    &    &    \\ 
\hline
12  &  1  &  18:30:09.69  &  -02:00:32.7  &  0.61 $\pm$   0.03  &  0.50 $\pm$   0.05  &  \multirow{5}{*}{GBS-VLA J183009.68-020032.7}  &  \multirow{5}{*}{Rad}  &  \multirow{5}{*}{6}  &    \\ 
12  &  2  &  --  &  --  &  $<$0.21  &  --  &    &    &    &    \\ 
12  &  3  &  --  &  --  &  $<$0.27  &  --  &    &    &    &    \\ 
12  &  4  &  --  &  --  &  $<$0.21  &  --  &    &    &    &    \\ 
12  &  all  &  18:30:09.68  &  -02:00:32.7  &  0.48 $\pm$   0.03  &  0.53 $\pm$   0.05  &    &    &    &    \\ 
\hline
13  &  1  &  18:30:25.88  &  -02:10:43.0  &  0.25 $\pm$   0.01  &  0.29 $\pm$   0.02  &  \multirow{5}{*}{2MASS J18302593-0210420, G028.5908+03.6734}  &  \multirow{5}{*}{Y*O}  &  \multirow{5}{*}{1}  &  \multirow{5}{*}{Y}  \\ 
13  &  2  &  18:30:25.88  &  -02:10:42.8  &  0.39 $\pm$   0.03  &  0.34 $\pm$   0.06  &    &    &    &    \\ 
13  &  3  &  --  &  --  &  $<$0.12  &  --  &    &    &    &    \\ 
13  &  4  &  18:30:25.88  &  -02:10:42.9  &  0.31 $\pm$   0.01  &  0.31 $\pm$   0.02  &    &    &    &    \\ 
13  &  all  &  18:30:25.88  &  -02:10:42.9  &  0.36 $\pm$   0.02  &  0.40 $\pm$   0.03  &    &    &    &    \\ 
\hline
14  &  1  &  --  &  --  &  $<$0.11  &  --  &    &    &    &  \multirow{5}{*}{Y}  \\ 
14  &  2  &  --  &  --  &  $<$0.09  &  --  &    &    &    &    \\ 
14  &  3  &  --  &  --  &  $<$0.17  &  --  &    &    &    &    \\ 
14  &  4  &  18:30:28.63  &  -01:53:32.6  &  0.23 $\pm$   0.01  &  0.21 $\pm$   0.02  &    &    &    &    \\ 
14  &  all  &  18:30:28.63  &  -01:53:32.7  &  0.19 $\pm$   0.01  &  0.16 $\pm$   0.02  &    &    &    &    \\ 
\hline
15  &  1  &  18:30:34.12  &  -01:56:37.5  &  0.60 $\pm$   0.04  &  0.67 $\pm$   0.09  &    &    &    &  \multirow{4}{*}{Y}  \\ 
15  &  2  &  --  &  --  &  $<$0.33  &  --  &    &    &    &    \\ 
15  &  4  &  --  &  --  &  $<$0.19  &  --  &    &    &    &    \\ 
15  &  all  &  18:30:34.12  &  -01:56:37.6  &  0.65 $\pm$   0.03  &  0.62 $\pm$   0.05  &    &    &    &    \\ 
\hline
16  &  1  &  --  &  --  &  $<$0.06  &  --  &  \multirow{5}{*}{IRAS 18280-0210, G028.6435+03.6457}  &  \multirow{5}{*}{Y*O}  &  \multirow{5}{*}{3}  &  \multirow{5}{*}{Y}  \\ 
16  &  2  &  --  &  --  &  $<$0.12  &  --  &    &    &    &    \\ 
16  &  3  &  18:30:37.61  &  -02:08:40.1  &  0.23 $\pm$   0.02  &  0.29 $\pm$   0.03  &    &    &    &    \\ 
16  &  4  &  --  &  --  &  $<$0.12  &  --  &    &    &    &    \\ 
16  &  all  &  18:30:37.61  &  -02:08:40.2  &  0.19 $\pm$   0.01  &  0.21 $\pm$   0.02  &    &    &    &    \\ 
\hline
17  &  1  &  --  &  --  &  $<$0.10  &  --  &    &    &    &  \multirow{5}{*}{Y}  \\ 
17  &  2  &  --  &  --  &  $<$0.08  &  --  &    &    &    &    \\ 
17  &  3  &  --  &  --  &  $<$0.13  &  --  &    &    &    &    \\ 
17  &  4  &  18:31:13.29  &  -02:05:46.2  &  0.65 $\pm$   0.02  &  0.63 $\pm$   0.03  &    &    &    &    \\ 
17  &  all  &  18:31:13.29  &  -02:05:46.2  &  0.22 $\pm$   0.01  &  0.17 $\pm$   0.02  &    &    &    &    \\ 
\hline 
\enddata
\tablenotetext{a}{Classification taken from the literature: young stellar object (Y*O), young stellar object candidate (Y*?), and known radio source (Rad). References: (1) \cite{Maury2011}; (2) \cite{Condon1998} (3); \cite{Winston2018AJ}; (4) \cite{Povich2013}; (5) \cite{Plunkett2018}; (6) \cite{OrtizLeon2015}}
\tablenotetext{b}{This flag indicates whether the source is a new radio continuum detection (Y) and/or has detected maser emission (H$_2$O). }
\end{deluxetable*}
\end{longrotatetable}
%
%\end{longtable}
%\tablefoot{
%\tablefoottext{a}{Classification taken from the literature: young stellar object (Y*O), young stellar object candidate (Y*?), and known radio source (Rad).} 
%References: (1) \cite{Maury2011}; (2) \cite{Condon1998} (3); \cite{Winston2018AJ}; (4) \cite{Povich2013}; (5) \cite{Plunkett2018}; (6) \cite{OrtizLeon2015}.
%}
%%\end{center}
%\end{landscape}
%%\end{longtab}

%-------------------------------------------------------------------

\appendix

\section{Supplementary Figures}\label{sec:appendix}

\begin{figure*}[!bht]
\begin{center}
{\includegraphics[width=0.99\textwidth,angle=0]{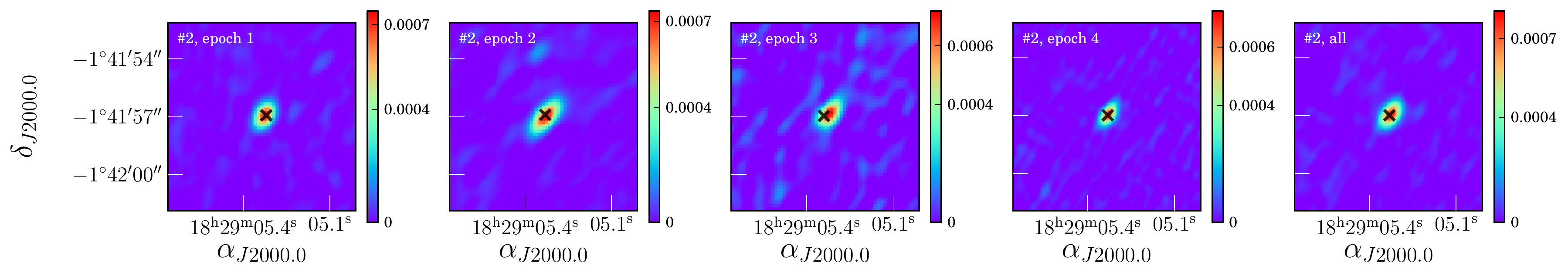}}
{\includegraphics[width=0.99\textwidth,angle=0]{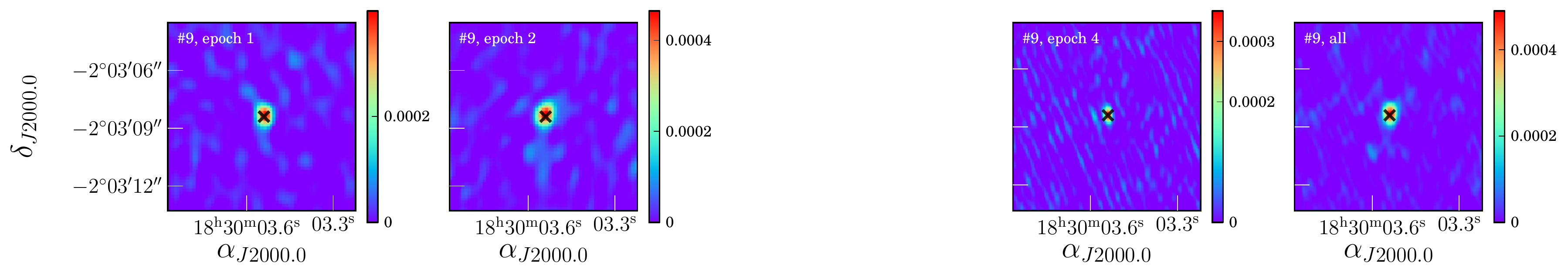}}
{\includegraphics[width=0.99\textwidth,angle=0]{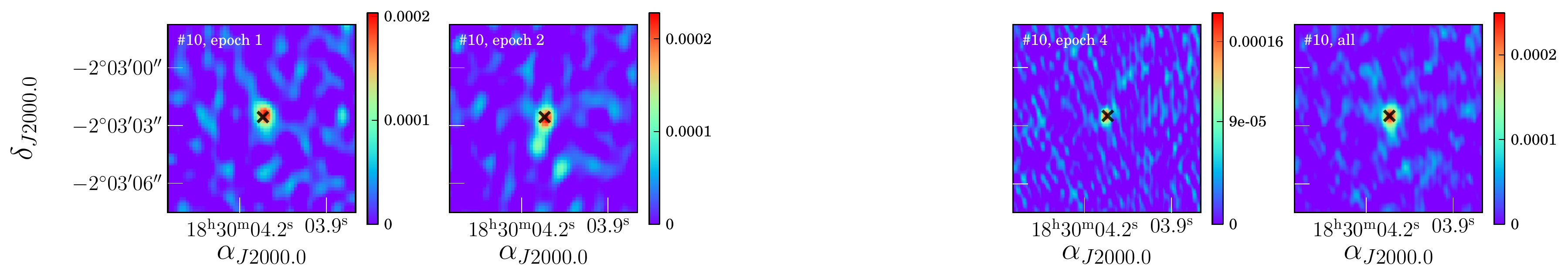}}
 \end{center}
\caption{Maps of radio continuum emission obtained with the VLA. The first 4 columns correspond to epochs 1, 2, 3 and 4,
 respectively (see Table \ref{tab:obs}). The last column shows the maps of the combination of epochs 1, 2 and 4. 
The color scale is in Jy~beam$^{-1}$. Shown are YSOs with detected water  masers, whose mean positions
are indicated by the black crosses.
}
\label{fig:vla-continuum-all}
\end{figure*}

\begin{figure*}[!bht]
\begin{center}
{\includegraphics[width=0.99\textwidth,angle=0]{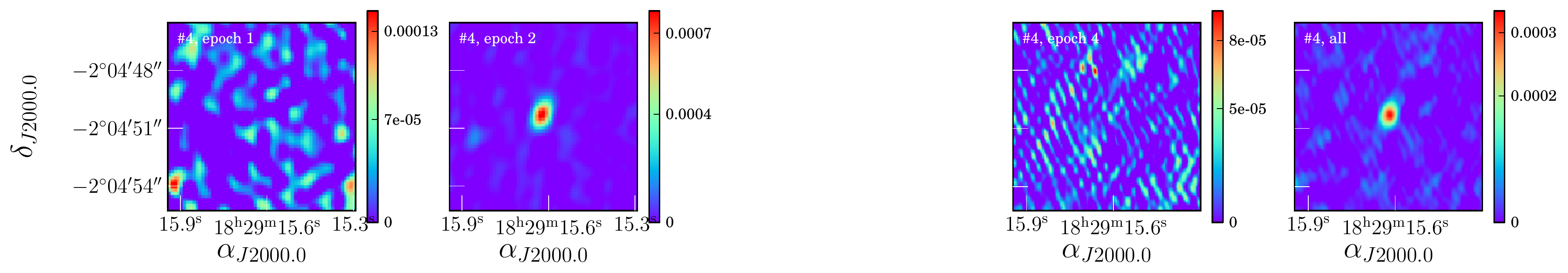}}
{\includegraphics[width=0.99\textwidth,angle=0]{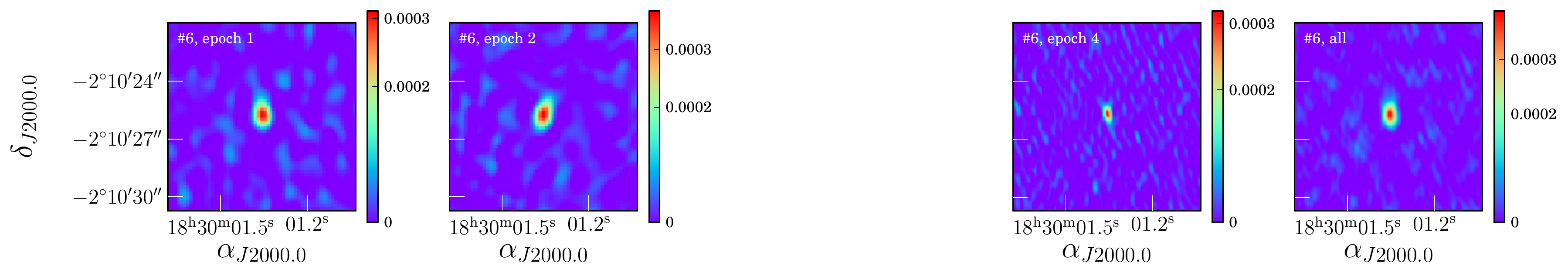}}
{\includegraphics[width=0.99\textwidth,angle=0]{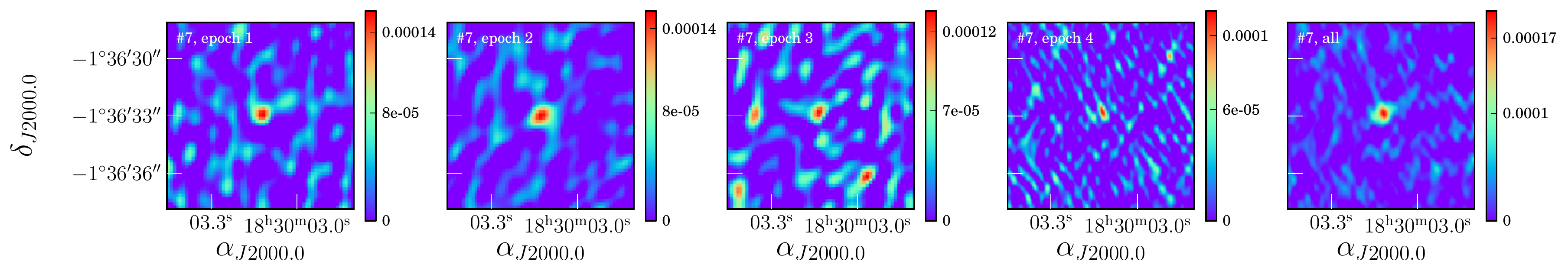}}
{\includegraphics[width=0.99\textwidth,angle=0]{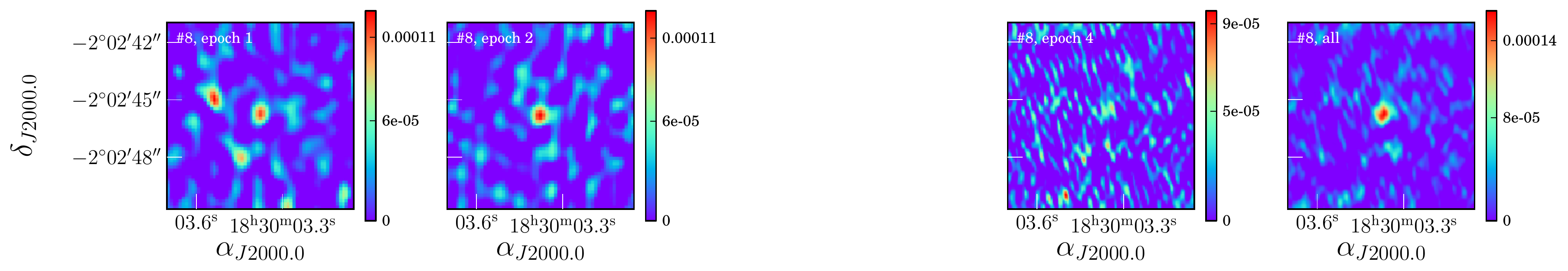}}
{\includegraphics[width=0.99\textwidth,angle=0]{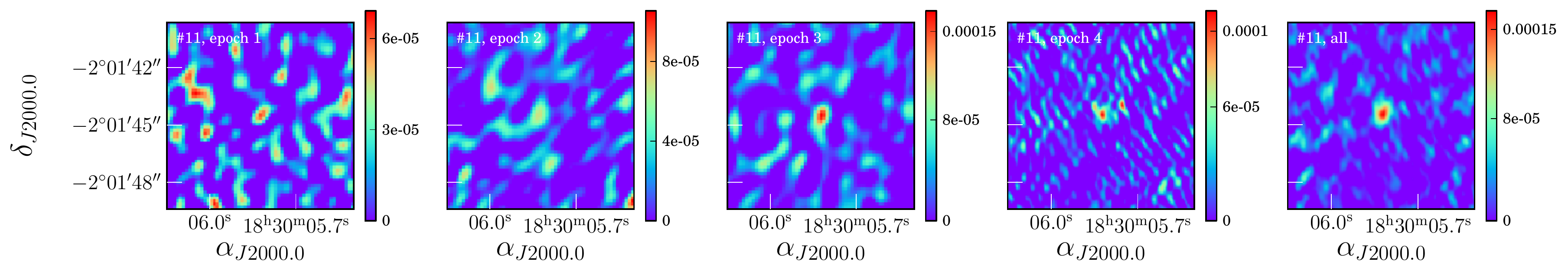}}
{\includegraphics[width=0.99\textwidth,angle=0]{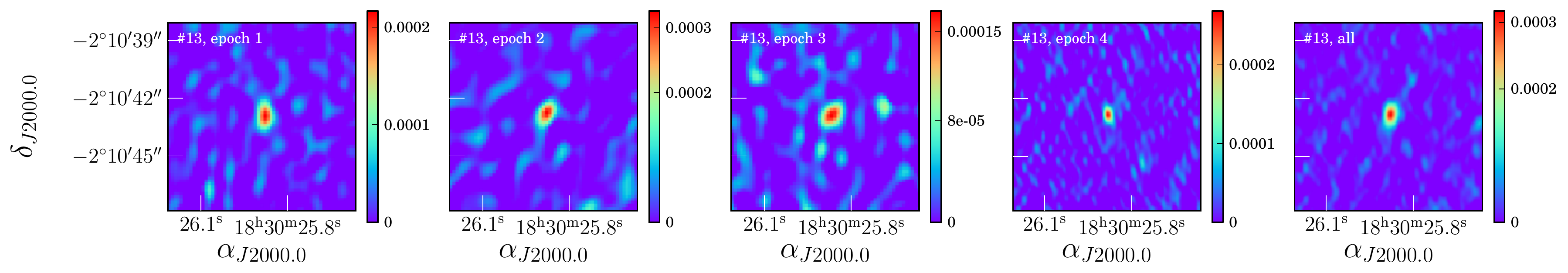}}
{\includegraphics[width=0.99\textwidth,angle=0]{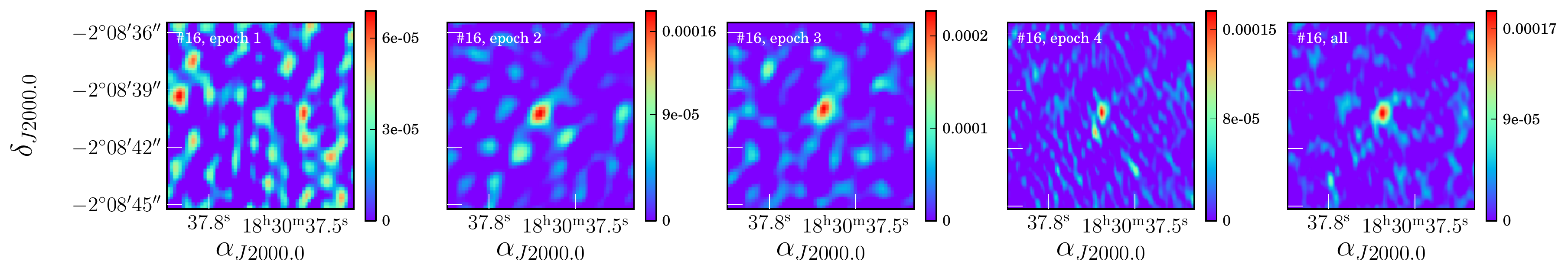}}
 \end{center}
\caption{As Fig. \ref{fig:vla-continuum-all}  for known or candidate YSOs and no detected water masers. }
\label{fig:vla-continuum-yso}
\end{figure*}

\begin{figure*}[bht]
\begin{center}
{\includegraphics[width=0.99\textwidth,angle=0]{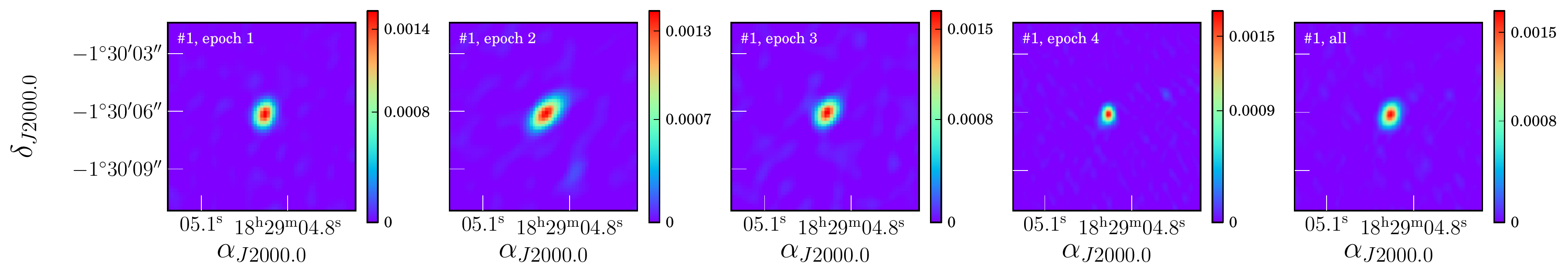}}
{\includegraphics[width=0.99\textwidth,angle=0]{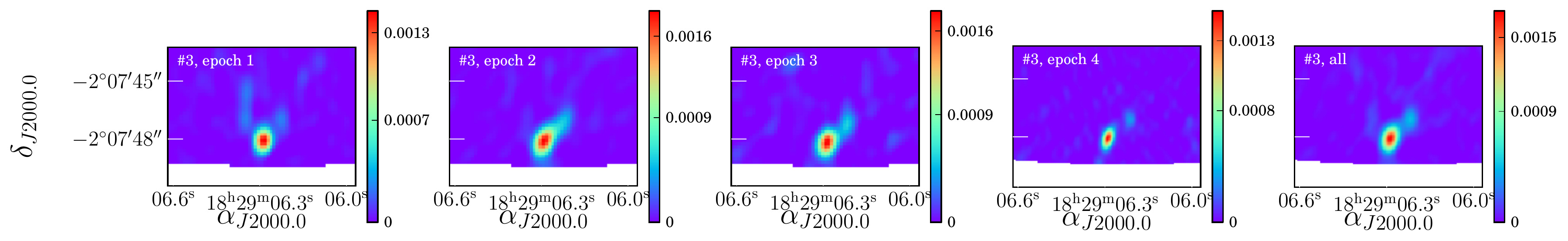}}
{\includegraphics[width=0.99\textwidth,angle=0]{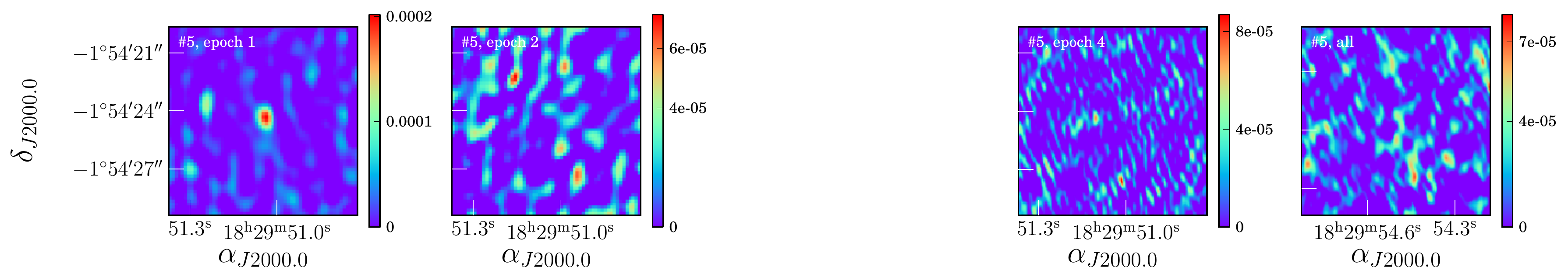}}
{\includegraphics[width=0.99\textwidth,angle=0]{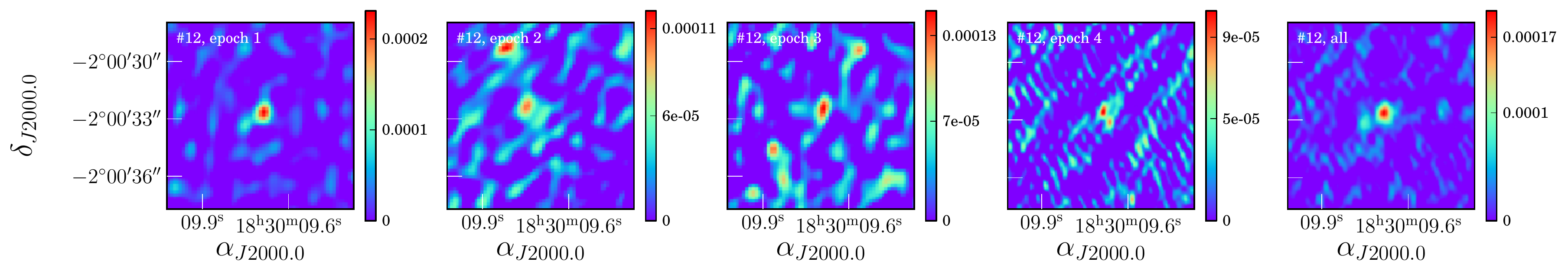}}
{\includegraphics[width=0.99\textwidth,angle=0]{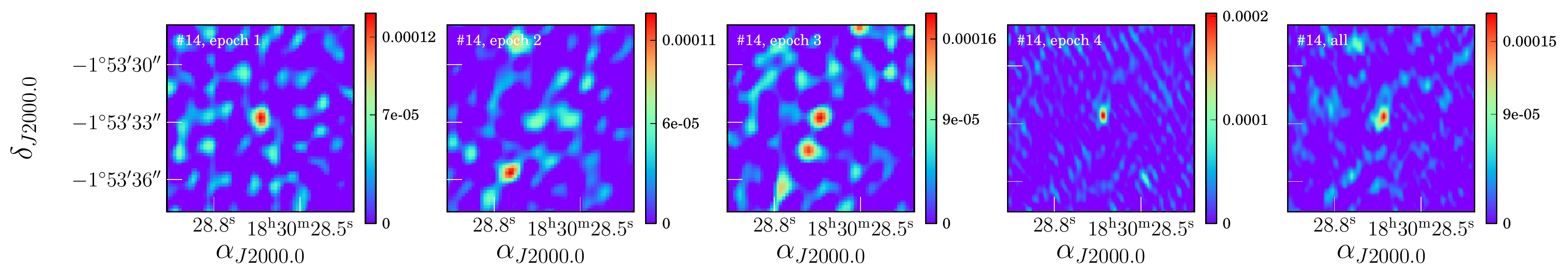}}
{\includegraphics[width=0.99\textwidth,angle=0]{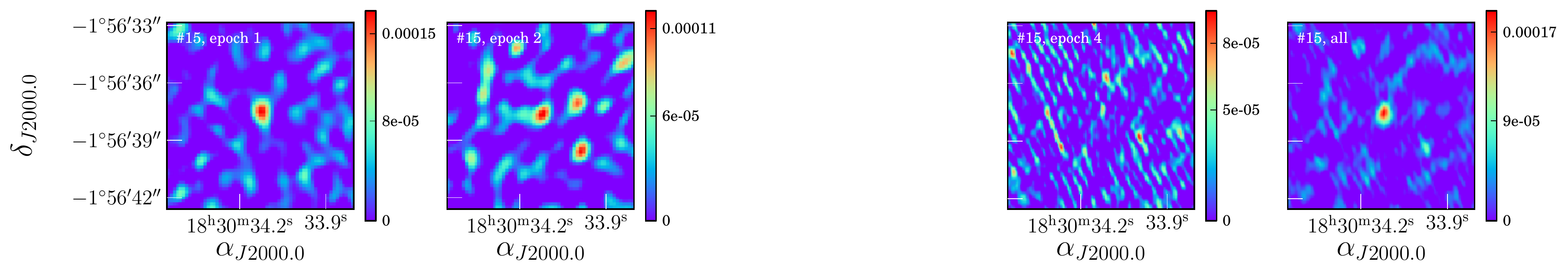}}
{\includegraphics[width=0.99\textwidth,angle=0]{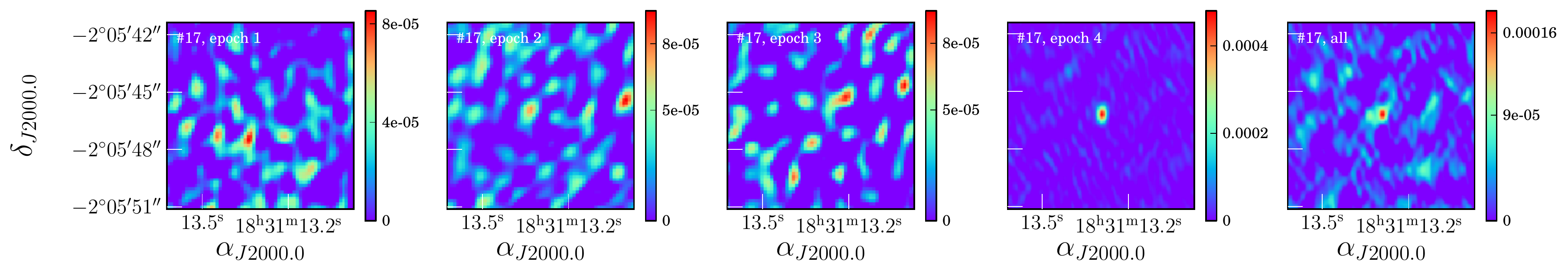}}
 \end{center}
\caption{As Fig. \ref{fig:vla-continuum-all}  for candidate extragalactic sources.}
\label{fig:vla-continuum-bck}
\end{figure*}

\begin{figure*}[bht]
\begin{center}
 {\includegraphics[width=0.5\textwidth,angle=0]{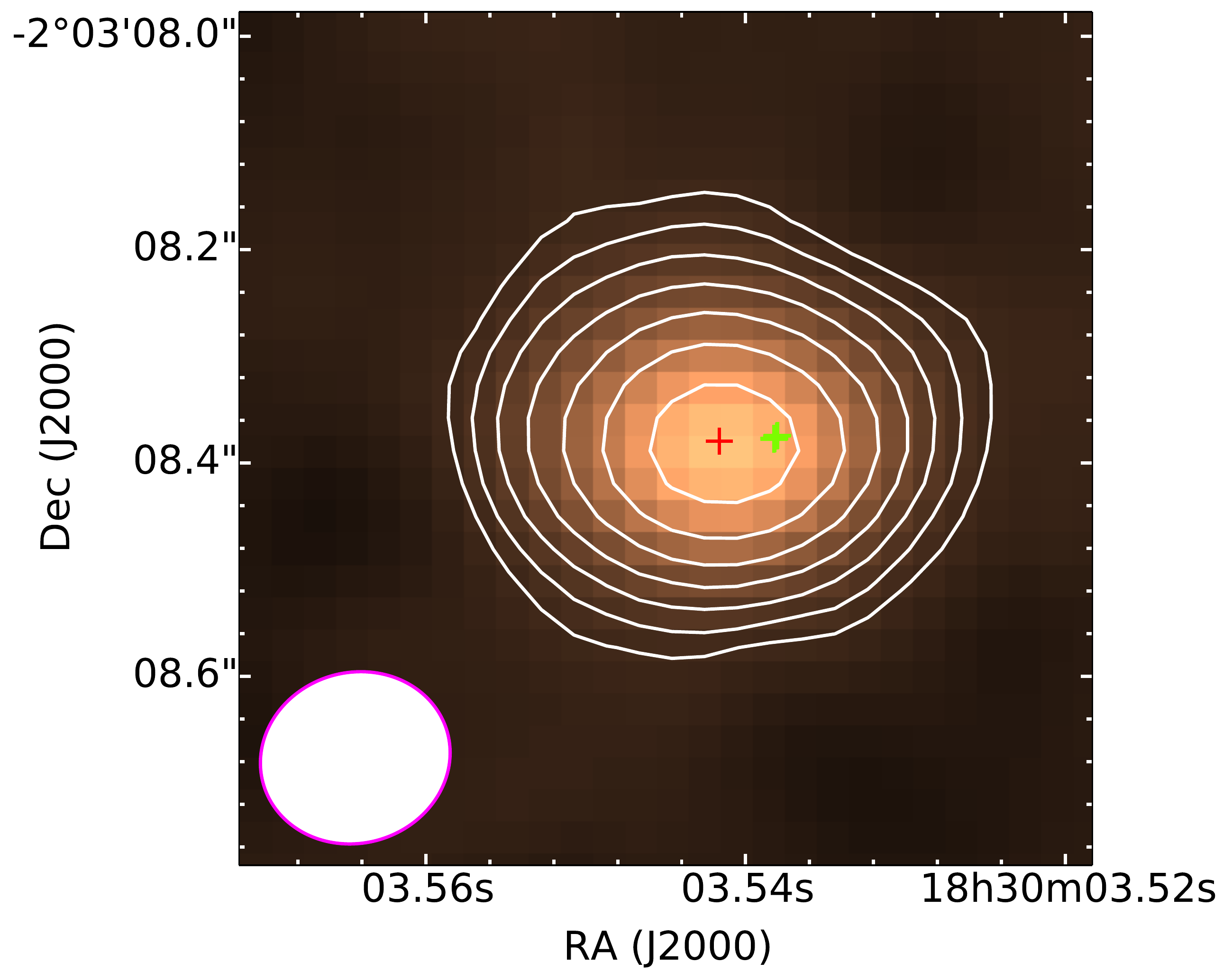}}
 \end{center}
\caption{ALMA continuum emission at 347~GHz from CARMA-6  (color scale and white contours).
 The $n$th contour is at $\left(\sqrt{2}\right)^{n}\times S_{\rm max} \times p$, where $S_{\rm max}=0.074$~Jy~beam$^{-1}$, $n$=0, 1, 2, 3, 4 ..., and $p$=10\%. 
 The green crosses indicate the positions of the water masers and the red cross correspond to the position of the ALMA continuum peak. The sizes of the crosses indicate 3 times the astrometric accuracy of ALMA  (9~mas) at 347~GHz.
 The beamsize is shown in white in the bottom left corner. 
 }
\label{fig:alma-cont-347}
\end{figure*}

In this Appendix, we show maps of radio continuum emission from the VLA data toward water maser sources (Figure \ref{fig:vla-continuum-all}), YSOs with no detected water masers (Figure \ref{fig:vla-continuum-yso}), and candidate extragalactic sources (Figure \ref{fig:vla-continuum-bck}). Figure \ref{fig:alma-cont-347} displays a 347 GHz ALMA continuum map of CARMA-6.

\bibliography{ms.bib}{}
\bibliographystyle{aasjournal}

%% This command is needed to show the entire author+affiliation list when
%% the collaboration and author truncation commands are used.  It has to
%% go at the end of the manuscript.
%\allauthors

%% Include this line if you are using the \added, \replaced, \deleted
%% commands to see a summary list of all changes at the end of the article.
%\listofchanges

\end{document}